\documentclass[aps,twocolumn,amsmath,amssymb,longbibliography,prb,10pt]{revtex4-2}
\usepackage[colorlinks=true,allcolors=blue,hypertexnames=false]{hyperref}
\usepackage{amsmath}
\usepackage{dsfont}
\usepackage{textcomp}
\usepackage{tabularx}
\usepackage{graphicx}
\usepackage{soul}
\usepackage{feynmp-auto}
\usepackage{footmisc}
\usepackage[export]{adjustbox}
\usepackage{wrapfig}
\usepackage{bbm}

\makeatletter
\def\@bibdataout@aps{%
 \immediate\write\@bibdataout{%
  @CONTROL{%
   apsrev42Control%
   \longbibliography@sw{%
    ,author="48",editor="1",pages="0",title="0",year="1"%
   }{%
    ,author="48",editor="1",pages="0",title="",year="1"%
   }%
  }%
 }%
 \if@filesw
  \immediate\write\@auxout{\string\citation{apsrev42Control}}%
 \fi
}%
\makeatother

\newcommand{\beg}{\begin{equation}}
\newcommand{\en}{\end{equation}}
\newcommand{\bp}{\mathbf p}
\newcommand{\bq}{\mathbf q}

\newcommand{\bk}{\mathbf k}
\newcommand{\br}{\mathbf r}

\newcommand{\bn}{\mathbf n}

\newcommand{\bE}{\mathbf E}

\newcommand{\ttaut}{\hat\tau_3}

\newcommand \bel  {\begin{align}}
\newcommand \enl  {\end{align}}

\newcommand{\eps}{\epsilon}
\newcommand{\veps}{\varepsilon}
\newcommand{\lam}{\lambda}

\newcommand{\up}{\uparrow}
\newcommand{\dn}{\downarrow}

\newcommand{\eref}[1]{Eq.~(\ref{#1})}

\renewcommand{\Re}{\mathrm{Re}}
\renewcommand{\Im}{\mathrm{Im}}
\newcommand{\Tr}{\mathrm{Tr}\,}

\newcommand{\ttau}[1]{\hat\tau_{#1}}
\newcommand{\itt}{i\hat\tau_2}
\newcommand{\gh}{\hat g}
\newcommand{\gc}{\check g}
\newcommand{\sw}{s\omega}
\newcommand{\hs}{\hat\sigma}
\newcommand{\dDh}{\delta\hat\Delta}
\newcommand{\etaR}{\eta^R_{\bn\eps}}
\newcommand{\etaA}{\eta^A_{\bn\eps}}
\newcommand{\ampl}{\left(\frac{ev_F}{\omega}\right)^{\!2}|\bn\!\cdot\!\bE|^2}
\renewcommand{\Re}{\mathrm{Re}}
\renewcommand{\Im}{\mathrm{Im}}

\newcommand{\gin}{\gamma_{\textrm{in}}}
\newcommand{\Gel}{\gamma_{\rm el}}
\newcommand{\tveps}{\tilde{\eps}}
\newcommand{\avg}[1]{\left\langle #1\right\rangle_\bn}
\newcommand{\Gcal}{{\cal G}}
\newcommand{\Fcal}{{\cal F}}
\newcommand{\Rcal}{{\cal R}}
\newcommand{\Ycal}{{\cal Y}}

\def\XXint#1#2#3{{\setbox0=\hbox{$#1{#2#3}{\int}$}
     \vcenter{\hbox{$#2#3$}}\kern-.5\wd0}}

\begin{document}

\title{Stimulation of superconductivity in $d$-wave superconductors}

\author{Aidan Steineman}
\affiliation{Department of Physics, Kent State University, Kent, Ohio 44242, USA}

\author{Maxim Dzero}
\affiliation{Department of Physics, Kent State University, Kent, Ohio 44242, USA}

\date{\today}

\begin{abstract}
We study the stimulation of superconductivity by an external electromagnetic radiation --- the
Eliashberg effect --- in a disordered $d$-wave superconductor. We use 
the self-consistent Keldysh--Nambu quasiclassical formalism augmented by an inelastic relaxation term
to compute the static correction to the pairing amplitude by treating elastic
scattering within the Born approximation with the full second-order
impurity vertex retained. We find that the clean $d$-wave superconductor shows
no enhancement at any frequency: without a momentum relaxation channel a spatially uniform drive produces no absorption,
only pair breaking. Elastic impurity scattering restores absorption and with it activates response in
the Eliashberg channel. We find enhancement of the pairing amplitude which is confined to a bounded
window in frequency, disorder strength, inelastic scattering rate and gap
magnitude.  Both edges of the window scale with the gap but are set by the competition
between quasiparticle redistribution, heating and pair breaking rather than by
a feature of the spectrum.  Furthermore, in contrast to the $s$-wave when the threshold is pinned at $2\Delta$, the upper edge is not pinned to the maximum gap and
crosses it as the system approaches the critical temperature.  
\end{abstract}

\maketitle
\section{Introduction}
Stimulated superconductivity --- the enhancement of the pairing
amplitude and of the superconducting critical temperature in conventional superconductors by a subgap electromagnetic drive --- is one of the
oldest and best established nonequilibrium effects in conventional
superconductors \cite{Dayem1967,Wyatt1966,Klapwijk1977,GM1986,KLAPWIJK2020,Kamenev2011}. A theory of this effect was proposed by Eliashberg \cite{Eliashberg1970} and then further developed in Refs. \cite{Ivlev1971,Ivlev1973}. The main idea is that the microwave radiation at frequencies $\omega <2\Delta$ cannot break Cooper pairs, but being absorbed by quasiparticles it leads to an excess of quasiparticles at higher energies and a deficit of quasiparticles at lower energies \cite{GM1986}. As a result, such a nonequilibrium state produces an increase in the superconducting order parameter and critical temperature. The effect persists as long as the redistributed quasiparticles relax slowly, i.e. the magnitude of the effect must be inversely proportional to the inelastic scattering rate, $\gamma_{\rm in}\ll \omega$. This phenomenon has been extensively discussed in the literature
\cite{KLAPWIJK2020} and continues to attract a lot of theoretical attention
\cite{TikhonovSkvortsovKlapwijk2018,Curtis2019,Eliashberg1,Eremin2024,Arora2025}.
 
Somewhat surprisingly, stimulation of superconductivity in $d$-wave
superconductors has received far less attention than for the $s$-wave case.
At the same time, light-induced superconductivity in the cuprates has been
the subject of intense study over the past decade
\cite{Fausti2011,Kaiser2014,Hu2014,Mankowsky2014,Cavalleri2018}, and since 
one may naively associate it with the effect considered here, we would like to highlight the
distinction right at the outset. Those experiments pump with femtosecond
mid-infrared pulses resonant with apical-oxygen phonons, at frequencies of
tens of THz that lie at or above the $d$-wave pair-breaking threshold $2\Delta_{\rm max}$ ($\Delta_{\rm max}=\sqrt{2}\Delta$ is the pairing gap in the antinodal direction and $\Delta$ is the pairing amplitude).
In contrast, the Eliashberg mechanism involves a subgap frequency by construction, and the enhancement window
found below lies lower still, at $\omega\simeq\Delta$.
 
In the light-induced superconductivity experiments the observable
is a transient $c$-axis Josephson plasma edge that appears for picoseconds
after the pulse. For example, in stripe-ordered
La$_{2-x}$Ba$_x$CuO$_4$ and La$_{1.675}$Eu$_{0.2}$Sr$_{0.125}$CuO$_4$ it
restores interlayer coherence suppressed by the stripe order
\cite{Fausti2011,Nicoletti2014}, while in underdoped
YBa$_2$Cu$_3$O$_{6+x}$ it appears above the equilibrium $T_c$
\cite{Kaiser2014,Hu2014}. The mechanisms proposed to explain
the pump--probe experimental observations --- nonlinear phononics \cite{Mankowsky2014},
parametric driving of the interlayer Josephson coupling
\cite{Denny2015,Hoppner2015}, phonon squeezing \cite{Kennes2017},
parametrically amplified electron--phonon coupling \cite{Sentef2016,Babadi2017}
--- all act on the pairing interaction or the phase stiffness at a fixed
quasiparticle distribution. In contrast with these what we compute here is the
static, second-order shift of the order parameter under a continuous
monochromatic drive at temperatures just below $T_c$, and the Eliashberg effect itself closes as $\Delta\to0$, so that there is no counterpart in the pseudogap regime $T\gtrsim T_c$. In addition, the Eliashberg effect originates from nonequilibrium quasiparticle distribution at
a fixed interaction. That body of theoretical work mentioned above therefore neither constrains nor is addressed by the
present calculation. What we predict here is a
distinct and so far unexplored effect: a steady-state enhancement of the
$d$-wave pairing gap under microwave or terahertz driving of a moderately
disordered superconductor at temperatures approaching the critical temperature $T_c$, with the drive at frequencies of the order of the gap amplitude,  
$\omega\simeq \Delta_{\rm max}/\sqrt{2}$.
 
\begin{figure*}
\includegraphics[width=\linewidth]{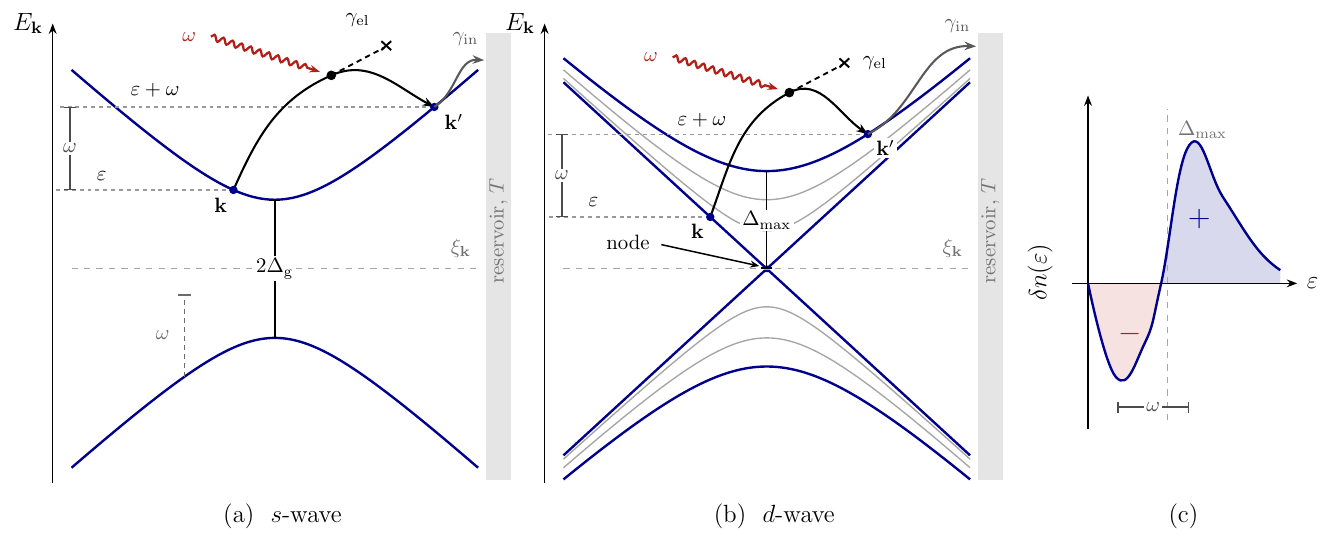}
\caption{Impurity-assisted absorption and the
nonequilibrium quasiparticle distribution it yields, which give rise to the Eliashberg effect. (a) $s$-wave case: the
quasiparticle branches $\pm E_\bk=\pm\sqrt{\xi_\bk^2+\Delta_{\rm g}^2}$ are
separated by $2\Delta_{\rm g}$. A thermally excited quasiparticle in the state
$\bk$ at energy $\eps$ absorbs a photon of energy $\omega$ (wavy line)
\emph{while} scattering off an impurity (dashed line, rate
$\gamma_{\rm el}$). The two vertices act as a single composite process in
which the field supplies the energy and the impurity supplies a channel for the momentum relaxation, so
that the final state $\bk'$ at energy $\eps+\omega$ is displaced along
$\xi_\bk$ by an amount of the order of $k_F$. The excited quasiparticle subsequently relaxes by
releasing its energy into the equilibrium reservoir at the inelastic scattering rate $\gamma_{\rm in}$ (grey
arrow). Direct pair breaking (vertical dashed arrow) is kinematically blocked
for $\omega<2\Delta_{\rm g}$. (b) $d$-wave case: with
$\Delta_\bn=\sqrt{2}\,\Delta\cos2\varphi$ the branches form a family (grey
lines illustrate intermediate directions on the Fermi surface) bounded by the antinodal
branch at $\Delta_{\rm max}=\sqrt{2}\Delta$ and degenerate at the node, so that
the same cycle operates at arbitrarily small $\eps$ and for any $\omega$.
Elastic scattering now changes the direction $\bn$ on the Fermi surface as well
as the magnitude of the momentum. (c) Resulting steady-state change of the
quasiparticle distribution: a deficit at low energies and a pile-up displaced
upward by $\omega$, shown relative to $\Delta_{\rm max}$ (vertical dashed
line). The dispersions in (a) and (b) are drawn to scale and the profile in (c) is
schematic, and its magnitude is proportional to $\gamma_{\rm in}^{-1}$.}
\label{Fig-Schematic}
\end{figure*}

In the case when microwave radiation is used to stimulate superconductivity there are three main physical ingredients which must be present for the Eliashberg effect to take place in a single band system. The first one is potential disorder which is needed for the light to be absorbed by the quasiparticles. Indeed, a quasiparticle in a state with momentum $\bk$ and energy $E_k$ can absorb a photon with energy $\omega$ only if its momentum changes from $\bk\to\bk'$. Otherwise the state with energy $E_\bk+\omega$ is not there and therefore no light absorption happens. Potential disorder provides such a momentum relaxation mechanism by unpinning the quasiparticle channel from the condensate, which in the case of conventional superconductors remains immune to potential scattering.
The same obstruction is well known for the nonlinear response: a
spatially uniform drive can be removed from a clean, Galilean-invariant
condensate by a boost, so that the coupling linear in the vector potential
excites neither the Higgs mode nor a third-harmonic signal unless impurity
scattering is present~\cite{Silaev2019}, a limitation that was noted to
persist for $d$-wave pairing~\cite{Steineman2026}. Also, keeping the photon momentum finite does not help because the corresponding quasiparticle energy transfer is $v_Fq_{\rm ph}=(v_F/c)\omega\ll \omega$. 
 
Since the elastic scattering cannot balance the upward energy flow of quasiparticles the second ingredient is needed: the inelastic scattering processes which are described by the inelastic scattering rate $\gamma_{\rm in}$. These processes anchor the quasiparticle distribution to a thermal bath and establish the time scale for a steady state with a nonequilibrium distribution function. Since by construction the change in the distribution function $\delta n(\eps)\propto \gamma_{\rm in}^{-1}$, inelastic scattering rate essentially determines the magnitude of the Eliashberg effect. 

Finally, the third equally important ingredient is the existence of the gap edge $\Delta_{\rm g}$ in conventional superconductors which not only guarantees that at frequencies $\omega<2\Delta_{\rm g}$ pair breaking processes will be kinematically forbidden but also allows for the quasiparticle re-distribution above the edge. Combining these ingredients, the overall physical picture of the Eliashberg effect reduces to: (i) at finite temperature thermally excited
quasiparticles occupy the states just above the edge, $\eps\gtrsim\Delta_{\rm g}$;
(ii) impurity-assisted absorption moves them to $\eps+\omega$, away from the
edge, at a rate proportional to $\Gel$ and to the intensity of the drive; (iii) if inelastic relaxation is slow they remain there for a time
$\gamma_{\rm in}^{-1}$. Since quasiparticles enter the self-consistency
equation with a weight that is largest at the gap edge, a distribution
depleted near $\Delta_{\rm g}$ suppresses the pairing amplitude less than the
thermal one, and $\delta\Delta>0$ for $\gamma_{\rm in}\ll\omega<2\Delta_{\rm g}$. One also finds an increased value of $T_c$ at temperatures $T\lesssim T_c$ provided the drive remains below $2\Delta(T)$ as the gap closes. In passing we note that while absorption requires momentum relaxation, of which disorder is one source while inelastic scattering may provide another, in the regime of interest the corresponding relaxation rates satisfy the following inequality: $\gamma_{\rm in}\ll \gamma_{\rm el}$. We illustrate this in Fig. \ref{Fig-Schematic}(a).
 
In conventional superconductors the condition $\omega<2\Delta$
guarantees that absorption leads to a redistribution of quasiparticles
at energies above the single particle excitation threshold only. This is clearly not so in the case of the disordered $d$-wave
superconductors where the existence of nodal quasiparticles provides the
absorption channel at any frequency. Furthermore, in the vicinity of the
nodes there is no gap to deplete: the excited nodal quasiparticles heat the
quasiparticle system while ending up in a steady state governed by inelastic
scattering. In the $d$-wave case potential disorder is also a pair
breaker. The Eliashberg effect in a $d$-wave superconductor is therefore a
competition between three processes: depletion of the gap edge in the
antinodal region, heating by nodal quasiparticles, and pair breaking by
impurities. The first two are enhanced by disorder and amplified by slow
inelastic relaxation, $\propto\gamma_{\rm in}^{-1}$, while the third grows
with disorder alone, Fig. \ref{Fig-Schematic}(b). The question is then which of them wins.
 
In this work we address precisely this question by employing the nonequilibrium quasiclassical
equations for $d$-wave superconductors in the presence of an external electromagnetic field to compute 
the second order static correction to the pairing amplitude. We found that whether the enhancement takes place or not is mainly determined by the dimensionless ratio $\omega/\Delta(T)$ provided that $\gamma_{\rm in}$ remains small. In this work we took $\gamma_{\rm in}\simeq 0.014\Delta_{00}$, for the elastic scattering rate we considered values of $\Gel$ in the interval $0.011\Delta_{00}\lesssim\gamma_{\rm el}\lesssim 0.40\Delta_{00}$ (here $\Delta_{00}$ is the pairing amplitude of the clean system at $T=0$). We find that for $\gamma_{\rm el}=0.072\Delta_{00}$ and the value of $\gamma_{\rm in}$ quoted above, for example, the enhancement region is bounded in frequency from both sides. Near its onset temperature, $T\simeq0.88\,T_c$, the region is narrow, $0.75\Delta(T)\lesssim\omega\lesssim1.33\Delta(T)$, and lies below $\Delta_{\rm max}$: at lower frequencies the nodal continuum dominates and $\delta\Delta<0$, while at higher frequencies pair breaking leads to the suppression of the pairing amplitude. Closer to $T_c$ the region widens on both sides, reaching $0.21\Delta(T)\lesssim\omega\lesssim2.16\Delta(T)$ at $T=0.98\,T_c$, so that its upper edge is not pinned to $\Delta_{\rm max}$ and thereby differs from the threshold condition $\omega\lesssim2\Delta_{\rm g}$ of the $s$-wave case. More generally, we find that unlike in the $s$-wave case, the boundary of the enhancement region in the $(T,\omega)$ plane is set by the competition between redistribution, heating and pair breaking in the superconducting kinetics and  not by the quasiparticle spectrum. We also find that within the validity of our perturbative approach, the increase in the values of the critical temperature is of order of several percent of the original $T_c$.
 
Our paper is organized as follows. Section~II sets up
the quasiclassical equation for the $d$-wave superconductor in an external electromagnetic field and its equilibrium limit. Section III is devoted to the perturbative analysis of the problem and the derivation of the expression for the static correction to the pairing amplitude $\delta\Delta$. Section~IV presents the results of our numerical calculation of $\delta\Delta$ for various values of the model parameters such as elastic and inelastic scattering rates as well as temperature. Section~V discusses the experimental implications of our results.
In Section VI we summarize the results of our work. 
Technical details are collected in the Appendices.

\section{Main equations}
\subsection{Model and formalism}
We consider a one-band model of 2D fermions with $d-$wave attractive interaction in the Cooper channel: 
\beg\label{Eq1}
\begin{aligned}
\hat{\cal H}&=\sum_{\bk,\sigma} \xi\left({\bk-\frac{e}{c}{\mathbf A}}\right) \hat{c}^\dagger_{\bk,\sigma}\hat{c}_{\bk,\sigma}\\&+
    \sum_{\bk,\bp,\bq} V_{\bk\bp} \hat{c}^\dagger_{\bk+\frac{\bq}{2},\uparrow} \hat{c}^\dagger_{-\bk+\frac{\bq}{2} ,\downarrow}\hat{c}_{-\bp+\frac{\bq}{2} ,\downarrow}\hat{c}_{\bp+\frac{\bq}{2},\uparrow},
\end{aligned}
\en
 where 
 $\hat{c}_{\bk\sigma}^\dagger$ ($\hat{c}_{\bk\sigma}$) are the creation (annihilation) fermionic operators, $\bk$ is momentum, $\sigma=\up,\dn$ is a spin,  
 $V_{\bk\bp}^{({\mathrm{d}})} $ is the pairing interaction,  $\xi(\bk)= k^2/2m - \veps_F$ is the single particle dispersion, and $\veps_F$ is the Fermi energy. 
     We project the interaction kernel $V_{\bk\bp}$ into the $d-$wave channel, $V_{\bk\bp}\to V_{\bk\bp}^{({\mathrm{d}})} $  and approximate it as  
$V_{\bk\bp}^{(\mathrm{d})}=- (\lambda/\nu_F){\cal Y}(\theta_\bk)\, {\cal Y}(\theta_\bp)$,
 where 
  $\lambda>0$ is the dimensionless coupling constant, $\nu_F$ is the density of states at the Fermi level, ${\cal Y}(\theta_\bk)=\sqrt{2}\cos2\theta_\bk$ is the normalized $d$-wave form factor and $\theta_\bk$ defines the direction of the momentum on the Fermi surface. A superconductor is subjected to an external electromagnetic radiation with the vector potential 
\beg\label{VectorPotential}
{\mathbf A}(\br,t)=\left(\frac{c{\mathbf E}}{i\omega}\right)e^{-i\omega t}+\left(-\frac{c{\mathbf E}^*}{i\omega}\right)e^{i\omega t}
\en
at zero photon momentum, $\bq=0$. 

In order to study the Elishberg effect we will use the standard Eilenberger--Keldysh
equations \cite{Eliashberg1972,LO1975,LO1977,SereneRainer1983,Kopnin2001}. We
summarize them here to fix the conventions. The main object in the subsequent discussion is the quasiclassical Green's function in Keldysh$\otimes$Nambu space defined according to
\beg\label{KeldyshProps}
\gc=\begin{pmatrix}\gh^R & \gh^K\\ 0 & \gh^A\end{pmatrix},
\en
which obeys the Eilenberger equation
\beg\label{EilenMain}
\begin{aligned}
\big[\veps\check\tau_3+\check\Delta_\bn-\check\Sigma\stackrel{\circ}{,}\gc\big]
&+\frac{i}{2}\big\{\check\tau_3,\partial_t\gc\big\}\\&
=-\frac{ev_F}{c}\big[(\bn\!\cdot\!{\mathbf A})\check\tau_3\stackrel{\circ}{,}\gc\big].
\end{aligned}
\en
Function $\check{g}$ satisfies the normalization condition 
\beg\label{Normg}
\gc\circ\gc=\check{\mathbbm 1}.
\en
The pairing field matrix is diagonal in Keldysh subspace 
$\hat\Delta_\bn=\itt\,\Ycal_\bn\Delta$ with the normalized $d$-wave form factor
$\Ycal_\bn=\sqrt2(n_x^2-n_y^2)$, $\avg{\Ycal_\bn^2}=1$, $\avg{\Ycal_\bn}=0$, where
\beg\label{FSaverage}
\avg{\cdots}=\int_0^{2\pi}\frac{d\phi_\bn}{2\pi}\{\cdots \}
\en
defines the Fermi-surface average. 
In \eqref{EilenMain} $X\circ Y$ denotes the
Groenewold--Moyal product, which for a component oscillating as $e^{-ia\omega t}$
reduces to the shift rule $(\hat X_a\circ\hat Y_b)(\eps)=\hat X_a(\eps+b\omega/2)\hat Y_b(\eps-a\omega/2)$.
The pairing amplitude must be determined self-consistently via
\beg\label{Self4Eli}
\Delta=\frac{\lam}{8}\int_{-\infty}^{\infty}\!d\eps\,
\avg{\Ycal_\bn\Tr\big[(-\itt)\,\hat{g}^K\big]}.
\en
Lastly, in \eqref{EilenMain} $\check{\Sigma}$ is the self-energy part with the same matrix structure as \eqref{KeldyshProps}. It is given by the sum $\check\Sigma=\check\sigma_{\rm in}+\check\sigma_{\rm dis}$ where $\check\sigma_{\rm in}$ is an 
inelastic part, which we take using in the reservoir model \cite{TikhonovSkvortsovKlapwijk2018}
\beg\label{sigma-in}
\check{\sigma}_{\rm in}=-\frac{i\gin}{2}\left(\begin{matrix} \ttau3 & 2t_\eps\,\ttau3 \\
0 & -\ttau3
\end{matrix}\right), 
\en
and we introduced the shorthand notation $t_\eps=\tanh({\eps}/{2T})$. Here $\gin$ is the rate at which the electronic distribution relaxes toward
the thermal one by exchange of energy (and quasiparticles) with the
environment---phonons leaving the film or a normal reservoir. It is
therefore set by the slowest step of that energy-escape chain. Collisions
between quasiparticles conserve the energy of the electronic system and do
not contribute to $\gin$ and we assume them slower than $\gin$, so that the
distribution computed below is the collisionless one. We also take a single
rate for energy and quasiparticle-number relaxation, which is the standard
reservoir (escape) approximation \cite{TikhonovSkvortsovKlapwijk2018}.

Consequently the elastic part due to scattering on point like  potential impurities is given by the following expression  
\beg\label{sigma-dis}
\check\sigma_{\rm dis}=-\frac{i}{2\tau}\avg{\gc},
\en
which is valid within the Born approximation and $\gamma_{\rm el}=\tau^{-1}$ is the elastic scattering rate. For the driven problem this is more than a convenience: solving the Eilenberger equation with the impurity collision integral in the presence of
the field was shown, for isotropic pairing, to be equivalent to the exact
summation of impurity-ladder diagrams with current vertices~\cite{Silaev2019}.

\subsection{Ground state}
Let us discuss the solution on \eqref{EilenMain} in equilibrium. We have 
\beg\label{ggsRA}
\gh^{R(A)}_{\bn\eps}=\ttau3\,g^{R(A)}_{\bn\eps}+\itt\,f^{R(A)}_{\bn\eps}
\en
and the Keldysh part is a simple parametrization $\gh^K_{\bn\eps}=\big(\gh^R_{\bn\eps}-\gh^A_{\bn\eps}\big)t_\eps$ \cite{Kamenev2009,Kamenev2011}.
The impurity self-energy in terms of these two averages
\beg\label{abg}
\avg{\gh^R_{\bn\eps}}=\ttau3\Gcal_\eps+\itt\Fcal_\eps. 
\en
For a $d$-wave order
parameter $\Fcal_\eps\equiv0$ identically: the equations of motion are covariant
under $\Delta_\bn\to-\Delta_\bn$, $\gc\to\ttau3\gc\ttau3$, so $f_{\bn\eps}$ is odd in
$\Ycal_\bn$ and averages to zero. Both self-energies then collapse into a single
complex energy variable which for retarded part is
\beg\label{xitilde}
\tveps^R_\eps=\eps+\frac{i\gin}{2}+\frac{i\gamma_{\rm el}}{2}\Gcal_\eps,
\en
and for the advanced part $\tveps^A_\eps=\big[\tveps^R_\eps\big]^*$ .
As a result, the propagators take the BCS form with $\Delta_\bn=\Ycal_\bn\Delta$,
\beg\label{gfeta}
g^{R(A)}_{\bn\eps}=\frac{\tveps^{R(A)}_\eps}{\eta^{R(A)}_{\bn\eps}},\qquad
f^{R(A)}_{\bn\eps}=\frac{\Delta_\bn}{\eta^{R(A)}_{\bn\eps}},
\en
so that $[g_{\bn\eps}^{R(A)}]^2-[f_{\bn\eps}^{R(A)}]^2=1$ on each branch. The denominators in \eref{gfeta} are defined according to 
\beg\label{etaRA}
\eta_{\bn\eps}^{R(A)}=\tveps^{R(A)}_\eps\sqrt{1-\left(\frac{\Delta_\bn}{\tveps^{R(A)}_\eps}\right)^2},
\en
with $\etaA=-\big[\etaR\big]^*$ and the analytic branch fixed by $\Im\,\etaR>0$.

\begin{figure}
\includegraphics[width=0.850\linewidth]{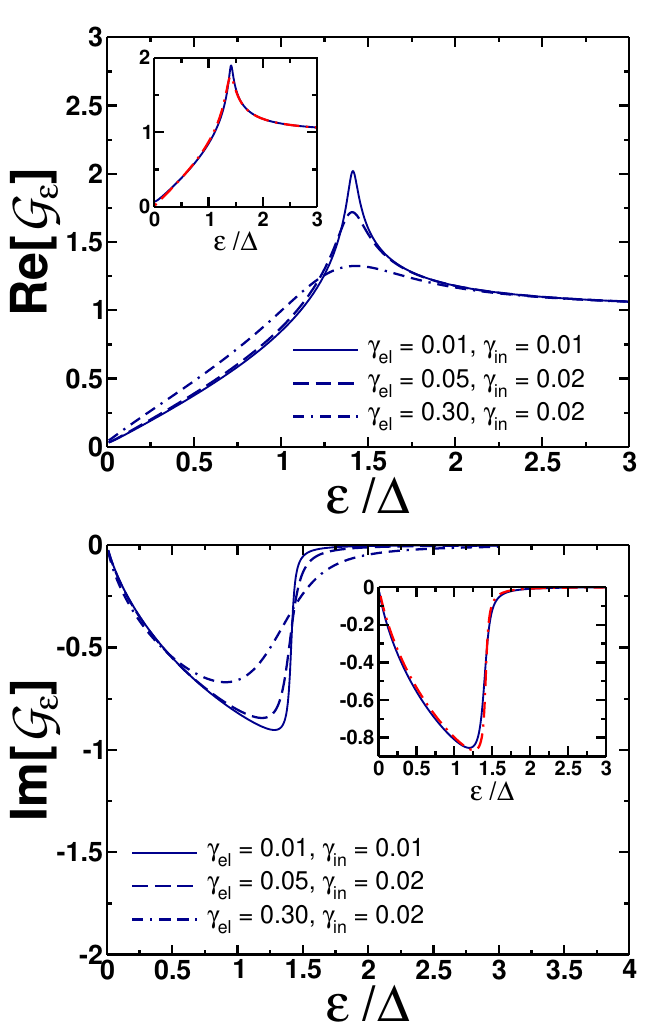}
\caption{Equilibrium impurity self-consistency of the $d$-wave state,
Eq.~(\ref{Gfixedpoint}): real (top panel) and imaginary (bottom panel)
parts of ${\cal G}_\eps=\avg{g^R_{\bn\eps}}$ as functions of $\eps/\Delta$
for different combination of values  $\gamma_{\rm el}$ and $\gamma_{\rm in}$ presented in the units of $\Delta$. Both
functions are shown for $\eps>0$ only, ${\rm Re}\,{\cal G}_\eps$ being even
and ${\rm Im}\,{\cal G}_\eps$ odd in $\eps$. ${\rm Re}\,{\cal G}_\eps$ is the
density of states normalized to its normal-state value: it is finite at
$\eps=0$ (the impurity-induced residual density of states of the nodal
superconductor), rises linearly through the nodal region, has its cusp at the
maximum gap $\Delta_{\rm max}=\sqrt2\Delta$, where the logarithmic
singularity of the clean $d$-wave state is smeared by the scattering rates,
and approaches unity for $\eps\gg\Delta_{\rm max}$. ${\rm Im}\,{\cal G}_\eps$
has the sign opposite to $\eps$ and enters the renormalized energy
$\tilde\eps^R_\eps$ of Eq.~(\ref{xitilde}) as a broadening: it is largest
just below $\Delta_{\rm max}$ and vanishes as $|\eps|\to\infty$. On the insets we compare ${\cal G}_\eps$ found self-consistently (solid line) 
with the corresponding expression with the constant broadening (dashed line).}
\label{Fig-calGeps}
\end{figure}

Inserting \eref{gfeta} into $\Gcal_\eps=\avg{g^R_{\bn\eps}}$ reduces the impurity problem to a closed equation for the single complex function ${\cal G}_\eps$. For the $d$-wave form factor it is given by 
\beg\label{Gfixedpoint}
\Gcal_\eps=\frac2\pi\,K\!\left(\frac{\sqrt2\Delta}{\tveps^R_\eps}\right),
\en
where 
$K(k)=\int_0^{\pi/2}d\vartheta/\sqrt{1-k^2\sin^2\vartheta}$ is a complete
elliptic integral, continued to complex
argument with the branch fixed by $\Gcal_\eps\to1$ as $|\eps|\to\infty$. The results of our self-consistent numerical solution of \eqref{Gfixedpoint} are shown in Fig. \ref{Fig-calGeps}. We emphasize that at each energy \eqref{Gfixedpoint} is a fixed-point equation for one complex number and the equations at different energies decouple which must be solved
rather than approximated by a constant broadening: the second-order Keldysh
response contains the denominators $\etaR+\etaA=2i\,\Im\,\etaR$, which are of order
$\gin$, and any inconsistency of order $\gamma_{\rm el}$ in the ground state is amplified by a factor
${\gamma_{\rm el}}/\gin$ (see insets in Fig. \ref{Fig-calGeps}). 

\begin{figure}
\includegraphics[width=0.850\linewidth]{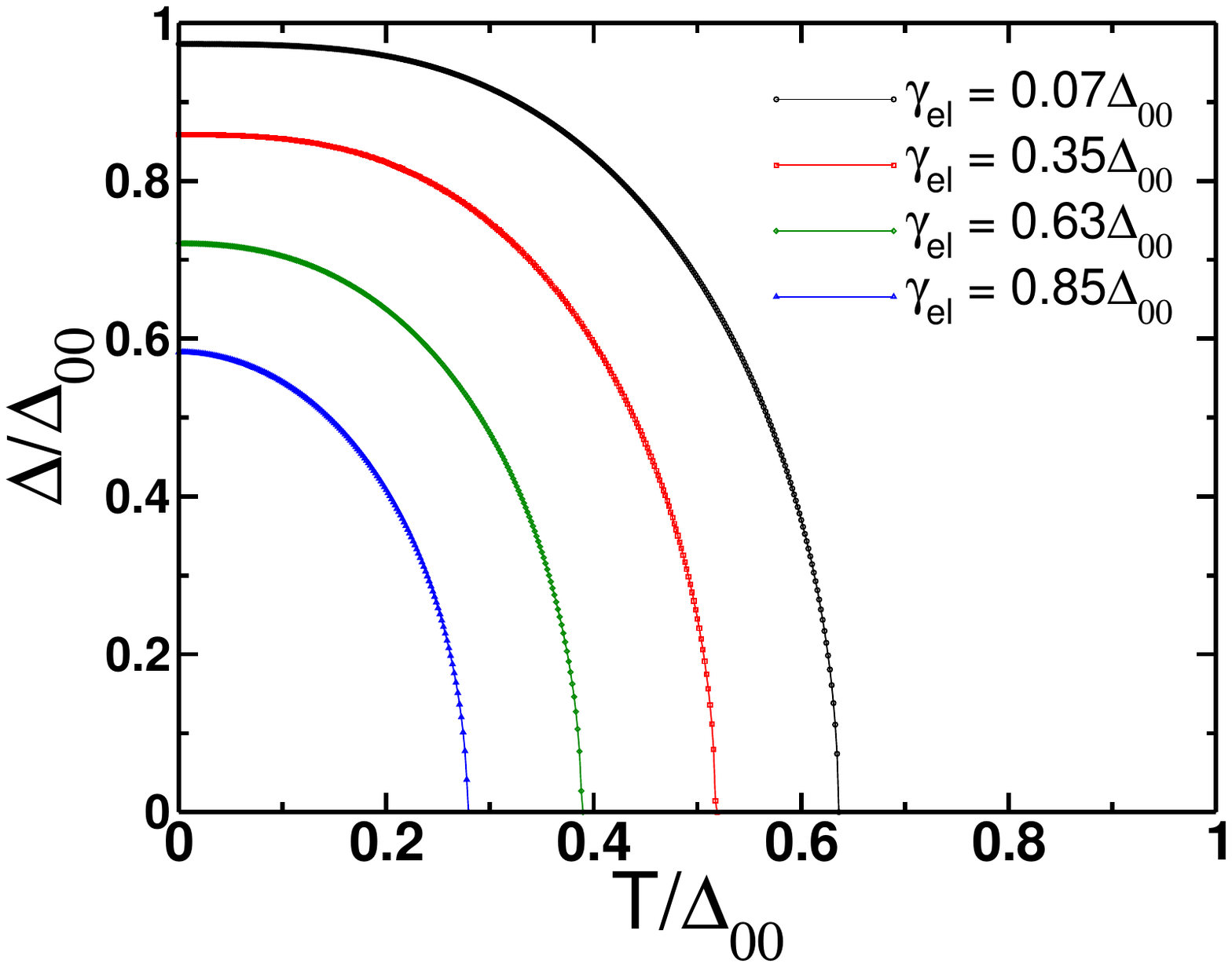}
\caption{Temperature dependence of the order parameter in the ground state found by solving the self-consistency equation \eqref{Self4Eli} for different values of the elastic scattering rate $\Gel$. Here $\Delta_{00}$ is the order parameter for a clean $d$-wave superconductor at $T=0$. }
\label{Fig-DLT}
\end{figure}

One can use the results for ${\Gcal_\eps}$ to compute the temperature dependence of the order parameter for different values of the elastic scattering rate. 
This dependence is shown in Fig. \ref{Fig-DLT}. As one might have expected, the order parameter amplitude is suppressed with increasing values of $\Gel$. We will use these results in what follows for the calculation of the nonlinear response functions. 

\section{Perturbative analysis}
In this Section we develop a perturbative approach which will allow us to derive an equation for the static correction to the pairing amplitude. The discussion here follows closely the derivation given in Ref. \cite{Steineman2026}. The basic idea is to expand $\gc=\gc_0+\gc_1+\gc_2+\dots$ in powers of the drive and then solve the Eilenberger equation at each order correspondingly. Given \eqref{VectorPotential} there are three contributions at second
order, each proportional to either $e^{\pm 2i\omega t}$ or $e^{i0\omega t}$ and we will keep the static component only. The static second-order correction
to the order parameter, $\dDh_\bn=\itt\,\Ycal_\bn\,\delta\Delta_\omega$, is then fixed by
the self-consistency condition
\beg\label{Self4Elig2K}
\delta\Delta_\omega=\frac{\lam}{2}\int_{-\infty}^{\infty}\!d\eps\,
\avg{\Ycal_\bn\Tr\big[(-\itt)\,\gh_2^K(\bn\eps;\omega)\big]}.
\en
The whole calculation therefore reduces to
finding the $\itt$ component of $\gh_2^K$, direction by direction, and projecting
it on $\Ycal_\bn$.

\subsection{First order corrections}\label{sec:first}
Writing $\gh_1=\gh_{1,+}e^{-i\omega t}+\gh_{1,-}e^{+i\omega t}$ and introducing the
drive amplitudes
\beg\label{amps}
{\cal A}_+=\frac{ev_F}{i\omega}(\bn\bE),\qquad {\cal A}_-=-\frac{ev_F}{i\omega}(\bn\bE^*),
\en
the retarded and advanced first-order corrections are (Appendix~\ref{app:first})
\beg\label{g1RA}
\gh^{R(A)}_{1,\pm}={\cal A}_\pm\,
\frac{\ttau3-\gh^{R(A)}_{\bn\eps\pm\frac\omega2}\ttau3\,\gh^{R(A)}_{\bn\eps\mp\frac\omega2}}
{\eta^{R(A)}_{\bn\eps+\frac\omega2}+\eta^{R(A)}_{\bn\eps-\frac\omega2}},
\en
For he Keldysh component we write $\gh^K_{1,\pm}=\gh^R_{1,\pm}\,t_{\eps\mp\frac\omega2}-\gh^A_{1,\pm}\,t_{\eps\pm\frac\omega2}
+\delta\gh^K_{1,\pm}$ with
\beg\label{g1K}
\begin{aligned}
&\delta\gh^K_{1,\pm}=
\frac{{\cal A}_\pm\big[\ttau3-\gh^{R}_{\bn\eps\pm\frac\omega2}\ttau3\,\gh^{A}_{\bn\eps\mp\frac\omega2}\big]
\big(t_{\eps\pm\frac\omega2}-t_{\eps\mp\frac\omega2}\big)}
{\eta^{R}_{\bn\eps\pm\frac\omega2}+\eta^{A}_{\bn\eps\mp\frac\omega2}} .
\end{aligned}
\en
These are the clean-limit expressions with $\eta_{\bn\eps}^{R(A)}$ are the functions of  $\tveps$ \eqref{xitilde}. The
reason is that the equilibrium part of the elastic self-energy
$-{i\gamma_{\rm el}}\Gcal_\eps\ttaut/2$ is time independent and proportional to
$\ttaut$, so that in the first-order equation its $\circ$-product with $\gh_1$
reduces to ordinary products with the
energy argument shifted by $\pm\omega/2$ exactly as for the term $\eps\ttaut$ alone. The three terms $\eps\ttaut$,
$({i\gamma_{\rm in}}/{2})\ttaut$ and $({i\gamma_{\rm el}}/{2})\Gcal_\eps\ttaut$ then combine into
$\tveps_{\eps\pm\omega/2}\ttaut$ and no separate impurity term survives. A
first-order correction to the self-energy itself, $\propto\avg{\gh_1}$, is
absent because $\gh_{1,\pm}\propto(\bn\cdot\bE)$ is odd under $\bn\to-\bn$.

All first-order blocks have the same matrix structure. With the notation
\beg\label{BCdef}
\begin{aligned}
&B_{\bn\eps}^{ab}\equiv g_{\bn\eps}^ag_{\bn\eps}^b+f_{\bn\eps}^af_{\bn\eps}^b-1,\\ 
&\overline{B}_{\bn\eps}^{ab}\equiv g_{\bn\eps}^ag_{\bn\eps}^b+f_{\bn\eps}^af_{\bn\eps}^b+1, \\
&C_{\bn\eps}^{ab}\equiv g_{\bn\eps}^af_{\bn\eps}^b+f_{\bn\eps}^ag_{\bn\eps}^b,
\end{aligned}
\en
one has
\beg\label{SimpleBCdef}
\gh_{\bn\eps}^a\ttau3\gh_{\bn\eps}^b-\ttau3=B_{\bn\eps}^{ab}\ttau3+iC_{\bn\eps}^{ab}\hat{\tau}_2,
\en
where $a,b$ stand for the two labels: energy with momentum direction ($\bn\eps$) and $X=R,A,K$ for the branch of the propagators involved.
Everything at second order is built from products and commutators of such blocks,
which is what makes the calculation tractable.

\subsection{Second order corrections: $\hat{g}_2^{R(A)}$}\label{sec:RA}
We continue with the calculation of the second order corrections to the retarded and advanced components of $\check{g}$. In contrast with the first order calculation, now have nonzero correction to the potential disorder part of self-energy $\hs_2\equiv-\frac{i}{2\tau}\avg{\gh_2}$, so that
equation for $\hat{g}_2^{R(A)}$ reads:
\beg\label{eq:g2RA}
\begin{aligned}
\big[\tveps^{R(A)}_\eps\ttau3+\hat\Delta_\bn,\gh_2^{R(A)}\big]
&=\hat{\cal R}^{R(A)}-\big[\dDh_\bn,\gh^{R(A)}_{\bn\eps}\big]\\&+\big[\hs_2^{R(A)},\gh^{R(A)}_{\bn\eps}\big], \\
\end{aligned}
\en
Functions $\gh_2^{R(A)}$ satisfy the following normalization condition
\beg\label{Norm4g2RA}
\gh^{R(A)}_{\bn\eps}\gh_2^{R(A)}+\gh_2^{R(A)}\gh^{R(A)}_{\bn\eps}=-\hat{\cal Z}^{R(A)},
\en
In \eqref{eq:g2RA} projector $\hat{\cal R}$ collects the drive acting on $\gh_1$ and $\hat{\cal Z}$ the
quadratic term of the normalization condition (the corresponding expressions for both are written out in
Appendix~\ref{app:RA}). The term $[\hs_2,\gh]$ is the second-order impurity vertex.
We emphasize that it plays an important role for it carries the conservation law of elastic scattering, and
dropping it replaces the $1/\gin$ relaxation of the isotropic distribution by
$1/\tau$. 

Since $\tveps\ttau3+\hat\Delta_\bn=\eta_{\bn\eps}\,\gh_{\bn\eps}$, any source of the
commutator form $[\hat X,\gh_{\bn\eps}]$ is solved by
$(\gh_{\bn\eps}\hat X\gh_{\bn\eps}-\hat X)/2\eta$, and the solution of \eref{eq:g2RA}
can now be easily written down as a sum $\hat{g}_2=\gh_2^{(0)}+\gh_2^{(1)}$:
\beg\label{g2RAsol}
\begin{aligned}
\gh_2&=\underbrace{\frac{\gh_{\bn\eps}\big(\hat{\cal R}-\eta_{\bn\eps}\hat{\cal Z}\big)}{2\eta_{\bn\eps}}
+\frac{\dDh_\bn-\gh_{\bn\eps}\dDh_\bn\gh_{\bn\eps}}{2\eta_{\bn\eps}}}_{\gh_2^{(0)}}\\
&+\underbrace{\frac{i\gamma_{\rm el}}{2}\,\frac{\avg{\gh_2}-\gh_{\bn\eps}\avg{\gh_2}\gh_{\bn\eps}}{2\eta_{\bn\eps}}}_{\gh_2^{(1)}}
\end{aligned}
\en
and we have suppressed the branch labels $R(A)$ for brevity. 
The drive part $\gh_2^{(0)}$ is known in closed form \cite{Steineman2026}. The vertex part $\gh_2^{(1)}$
depends on the Fermi-surface average $\avg{\gh_2}$, which remains to be found so that equation \eref{g2RAsol} is
an integral equation in $\bn$. 

Equation \eqref{g2RAsol} can be most easily solved by projection. We will look for solution in the following general form:
\beg\label{avg2ansatz}
\langle{\hat g_2^{R(A)}}\rangle_\bn=X_0\hat\tau_0+X_1\hat\tau_1+i\hat\tau_2X_2
+X_3\hat\tau_3,
\en
where we suppressed the branch labels for brevity. 
The pre-factor in front of $\gamma_{\rm el}$ in the expression for $\hat{g}_2^{(1)}$ can be interpreted as the disorder vertex renormalization, so we introduce ${\mathcal V}[\hat{G}]=(\hat{g}_{\bn\eps}\hat{G}\hat{g}_{\bn\eps}-\hat{G})/2\eta_{\bn\eps}$. Given \eqref{Normg} it follows 
that ${\mathcal V}[\hat{\tau}_0]=0$ which implies that $X_0=0$. Next we consider ${\mathcal V}[\hat{\tau}_1]$ and using \eqref{Normg} we have 
${\mathcal V}[\hat{\tau}_1]=-\hat{\tau}_1/\eta_{\bn\eps}$ so we find $X_1=\langle{\textrm{Tr}\big[\ttau1\gh_2^{(0)}\big]}\rangle_\bn/{2\left[1-i\gamma_{\rm el}\avg{\eta_{\bn\eps}^{-1}}/2\right]}=0$ since $\gh_2^{(0)}$ has no $\ttau0$ or $\ttau1$ components 
(Appendix~\ref{app:RA}). Note that ${\mathcal V}$ mixes the $\ttau3$ and $\itt$ channels,
${\mathcal V}[\ttau3]=(f^2\ttau3+gf\,\itt)/\eta_{\bn\eps}$ and
${\mathcal V}[\itt]=-[(1+f^2)\itt+gf\,\ttau3]/\eta_{\bn\eps}$, so that $X_2$ and
$X_3$ satisfy a coupled pair of equations in general. For $d$-wave pairing the
coupling between $X_2$ and $X_3$vanishes by $d$-wave symmetry: $\langle{g^{R(A)}_{\bn\eps}f^{R(A)}_{\bn\eps}/\eta^{R(A)}_{\bn\eps}}\rangle_\bn=0$. As a result the two coefficients decouple:
\beg\label{X3closure}
\begin{aligned}
X_3^{R(A)}
&=\frac{\avg{\textrm{Tr}\big[\hat{\tau}_3\gh_2^{(0)}\big]^{R(A)}}}{2\left[1-\gamma_{\rm el}\avg{\frac{(f^{R(A)}_{\bn\eps})^2}{2i\eta^{R(A)}_{\bn\eps}}}\right]}, \\
X_2^{R(A)}&=-\frac{\avg{\textrm{Tr}\big[i\hat{\tau}_2\gh_2^{(0)}\big]^{R(A)}}}{2\left[1+\gamma_{\rm el}\avg{\frac{(g^{R(A)}_{\bn\eps})^2}{2i\eta^{R(A)}_{\bn\eps}}}\right]}.
\end{aligned}
\en
We can now insert these expressions into \eqref{g2RAsol} which gives us the closed form expressions for the functions $\hat{g}_2^{R(A)}$. 

\subsection{Second order corrections: $\hat{g}_2^{K}$}\label{sec:K}
As in the linear analysis we use the following ansatz for the Keldysh component:
\beg\label{g2Kansatz}
\gh_2^K=\big(\gh_2^R-\gh_2^A\big)t_\eps+\delta\gh_2^K ,
\en
with the \emph{full} $\gh_2^{R,A}$ of \eref{g2RAsol}, vertex part included. The
anomalous part splits like the spectral one, $\delta\gh_2^K=\delta\gh_2^{K,0}+\delta\gh_2^{K,1}$.
The drive part is (Appendix~\ref{app:K})
\beg\label{dg2K0}
\delta\gh_2^{K,0}(\bn\eps;\omega)=\frac{\gh^R_{\bn\eps}\,\hat{\cal S}(\bn\eps;\omega)}{\etaR+\etaA},
\en
where $\hat{\cal S}\equiv\hat{\cal R}^K+\hat{\cal R}^{\delta K}-\etaA\hat{\cal D}=S_0\ttau0+S_1\ttau1$ and functions $\hat{\cal R}^K$, $\hat{\cal R}^{\delta K}$, $\hat{\cal D}$ are built from the
first-order blocks (\ref{g1RA})--(\ref{g1K}) and are given, together with the
scalars $S_0$, $S_1$, in Appendix~\ref{app:K}. The source $\hat{\cal S}$ has only
$\ttau0$ and $\ttau1$ components, because every term in it is a product of two
matrices from the span of $\{\ttau3,\itt\}$ --- a first-order block \eref{BCdef}
times $\ttau3$, or two such blocks. Consequently
\beg\label{dg2K0comp}
\begin{aligned}
\frac{1}{2i}\textrm{Tr}\big[\hat{\tau}_2\delta\gh_2^{K,0}\big]=\frac{g^R_{\bn\eps}S_1+f^R_{\bn\eps}S_0}{\etaR+\etaA}.
\end{aligned}
\en 

Let us now discuss the vertex part. The vertex enters the Keldysh equation through
$-\frac{i}{2\tau}[\avg{\gc_2},\gc]^K$, whose four terms rearrange, with
\eref{g2Kansatz}, into
\beg\label{Kcomm}
\begin{aligned}
\big[\avg{\gc_2},\gc\big]^K
&=\Big(\big[\avg{\gh_2^R},\gh^R\big]-\big[\avg{\gh_2^A},\gh^A\big]\Big)t_\eps
\\&+\avg{\delta\gh_2^K}\gh^A-\gh^R\avg{\delta\gh_2^K} .
\end{aligned}
\en
The first bracket is exactly the impurity source of the retarded and advanced
equations multiplied by $t_\eps$, and it is consumed by the
$(\gh_2^R-\gh_2^A)t_\eps$ part of the ansatz: for any pair $\gh_2^{R,A}$ satisfying
their own equations, the Keldysh operator acting on $(\gh_2^R-\gh_2^A)t_\eps$
reproduces $t_\eps$ times the difference of the $R$ and $A$ equations, the
mismatch $(\tveps^R-\tveps^A)$ being supplied by $\hs^K_0$
(Appendix~\ref{app:K}). The anomalous vertex part is therefore sourced by the
last two terms of \eref{Kcomm} alone, and by the same particular-solution formula
\beg\label{dg2K1}
\delta\gh_2^{K,1}=\left(-\frac{i\gamma_{\rm el}}{2}\right)\,\frac{\gh^R_{\bn\eps}\avg{\delta\gh_2^K}\gh^A_{\bn\eps}-\avg{\delta\gh_2^K}}{\etaR+\etaA}.
\en
Writing $\avg{\delta\gh_2^K}=Y_2\,\itt+ Y_3\ttau3$, the two components needed are
\beg\label{KSystem}
\begin{aligned}
\frac{1}{2i}\textrm{Tr}\big[\hat{\tau}_2(\gh_{\bn\eps}^R\avg{\delta\gh_2^K}\gh_{\bn\eps}^A&-\avg{\delta\gh_2^K})\big]\\&
=Y_3C_{\bn\eps}^{RA}-Y_2\overline{B}_{\bn\eps}^{RA},\\
\frac{1}{2}\textrm{Tr}\big[\hat{\tau}_3(\gh_{\bn\eps}^R\avg{\delta\gh_2^K}\gh_{\bn\eps}^A&-\avg{\delta\gh_2^K})\big]\\&=Y_3B_{\bn\eps}^{RA}-Y_2C_{\bn\eps}^{RA}.
\end{aligned}
\en
We now average both parts of each equation with respect to $\bn$. 
The cross coefficients $C_{\bn\eps}^{RA}$, Eq. \eqref{BCdef}, are odd in
$\Ycal_\bn$ and average to zero, so $Y_2$ and $Y_3$ decouple, and
$Y_2$ drops out of the $\Ycal_\bn$-projection for the same reason as $X_2$.
Averaging the $\ttau3$ component of $\delta\gh_2^{K,0}+\delta\gh_2^{K,1}$ gives
the following equation for $Y_3$:
\beg\label{dY3closure}
\begin{aligned}
&Y_3\left[1+\frac{i\gamma_{\rm el}}{2}\avg{\frac{B_{\bn\eps}^{RA}}{\etaR+\etaA}}\right]
=\avg{\frac{g^R_{\bn\eps}S_0+f^R_{\bn\eps}S_1}{\etaR+\etaA}} .
\end{aligned}
\en
In the limit $T\to T_c$, where $\Delta\to0$, $g^R_{\bn\eps}g^A_{\bn\eps}\to-1$
and $\etaR+\etaA\to i(\gin+\Gel)$, both sides of \eref{dY3closure} carry the
total rate: the source through $(\etaR+\etaA)^{-1}$, and the bracket, which
reduces to $\gin/(\gin+\Gel)$. In $Y_3$ the combination $\gin+\Gel$ therefore
cancels out, leaving $\gin^{-1}$. This is not special to the limit: elastic
scattering only redistributes quasiparticles over the Fermi surface at fixed
energy, and therefore cannot set the scale of the response of the
Fermi-surface average at a given $\eps$. At finite $\Delta$ the bracket still
vanishes linearly in $\gin$, so that $Y_3\propto\gin^{-1}$ at every energy.
The elastic rate enters the spectral functions and the $O(1)$ coefficient, but
not the enhancement factor. This is the entire disorder-assisted redistribution effect, and it is lost if
the Keldysh impurity vertex $\delta\gh_2^{K,1}$ of \eref{dg2K1} is dropped:
the bracket in \eref{dY3closure} then reduces to unity and $Y_3(\eps)$ is governed
by $(\gin+\Gel)^{-1}\approx \Gel^{-1}$ instead of $\gin^{-1}$, a suppression by a factor of
$(1+\Gel/\gin)$.

\begin{figure}
\includegraphics[width=0.850\linewidth]{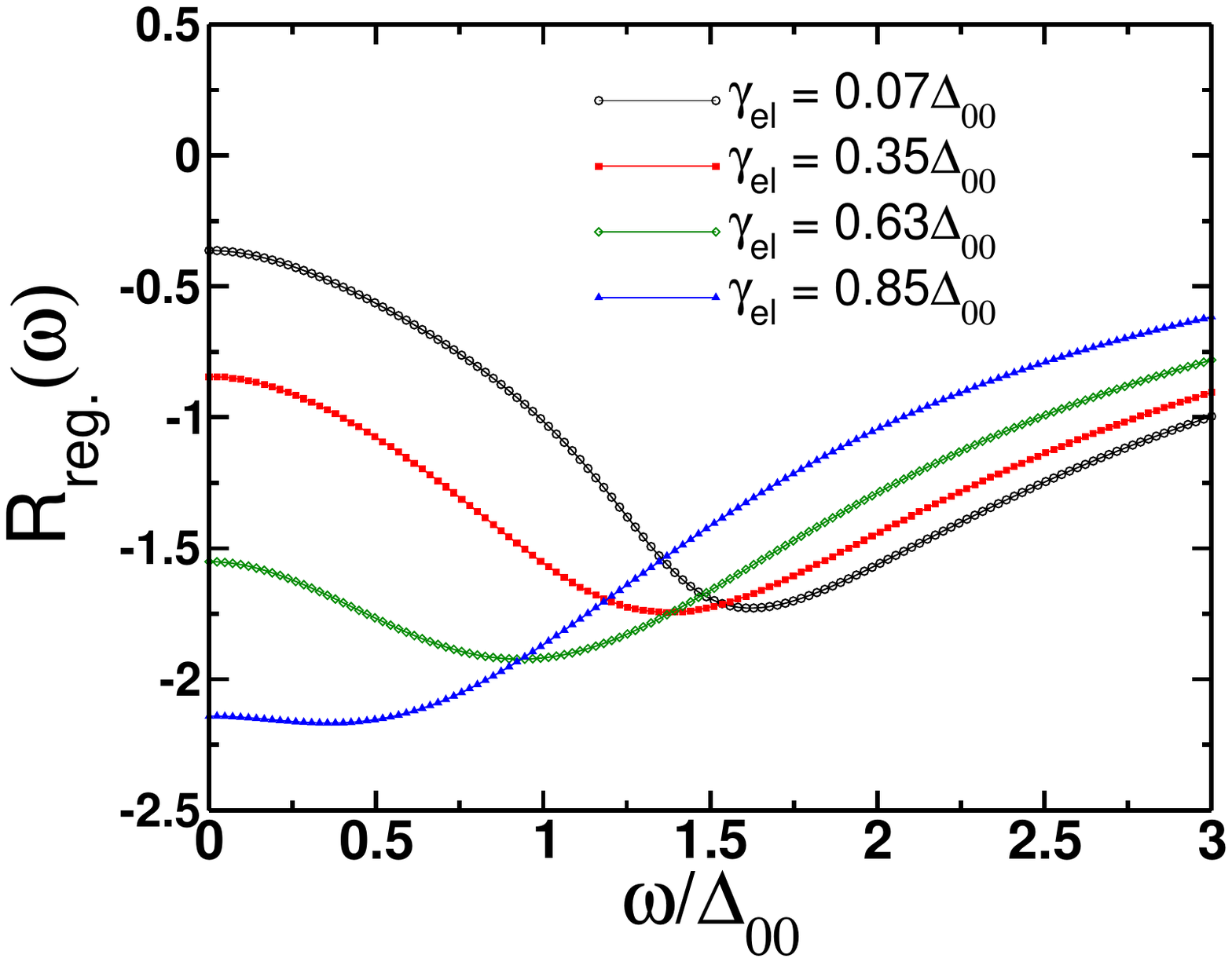}
\caption{Frequency dependence of the regular part of the photo-kernel $\Rcal_{\rm reg}(\omega)$, Eq. \eqref{R-kernel}, for different values of the elastic scattering rate. For this plot we also used $\gamma_{\rm in}=0.0144\Delta_{00}$ and $T=0.2\Delta_{00}$ where $\Delta_{00}$ is the value of the $d$-wave superconducting order parameter for a clean system at $T=0$.} 
\label{Fig-Rreg}
\end{figure}

\begin{figure}
\includegraphics[width=0.850\linewidth]{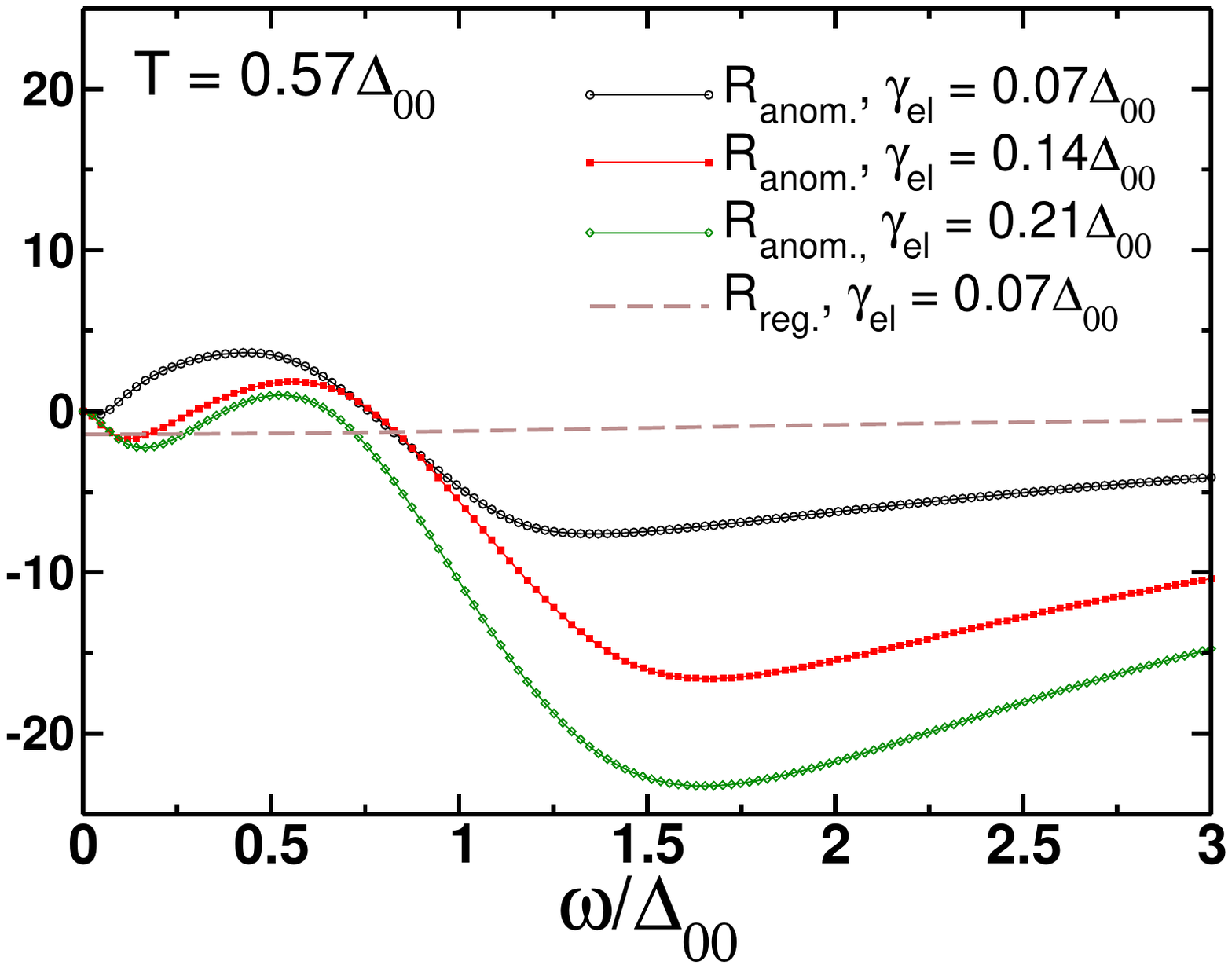}
\caption{Frequency dependence of the anomalous part of the photo-kernel $\Rcal_{\rm anom.}(\omega)$, Eq. \eqref{R-kernel}, for different values of the elastic scattering rate. For this plot we also used $\gamma_{\rm in}=0.0144\Delta_{00}$ where $\Delta_{00}$ is the value of the $d$-wave superconducting order parameter for a clean system at $T=0$. Since $R_{\rm reg.}(\omega)$ shows very weak dependence for the chosen range of parameters, here we show its frequency dependence for only one value of $\Gel$ for comparison.} 
\label{Fig-Ranom}
\end{figure}

\begin{figure*}
\includegraphics[width=0.95\linewidth]{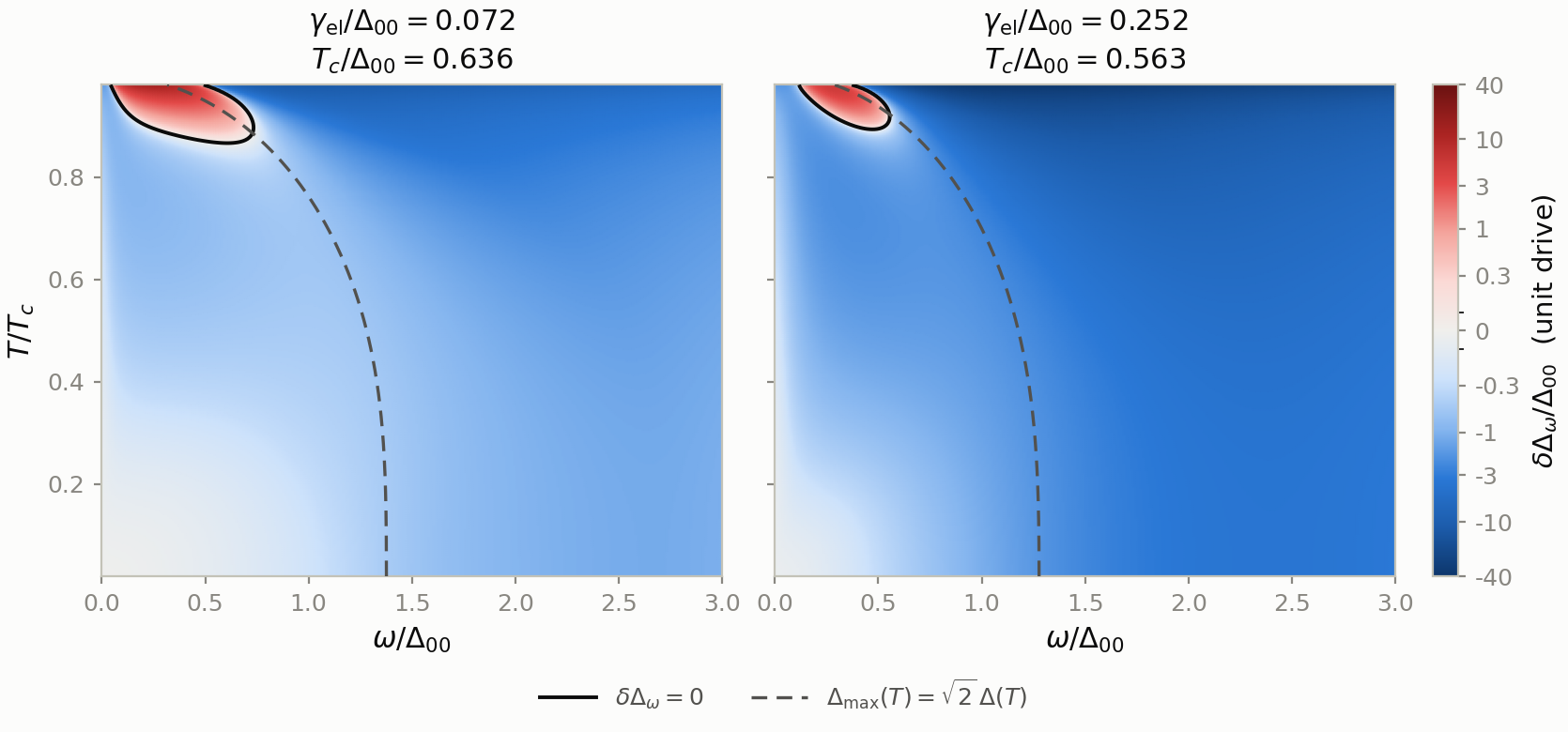}
\caption{Photo-induced change of the $d$-wave order parameter,
$\delta\Delta_\omega/\Delta_{00}$, on the $(\omega,T)$ plane for two elastic
scattering rates, $\Gel/\Delta_{00}=0.072$ (left) and $0.252$ (right), with
$T_c/\Delta_{00}=0.636$ and $0.563$ respectively.  The quantity shown is the
coefficient of $(ev_FE/\omega)^2$ in Eq.~\eqref{Form4Eli}, i.e.\
$\delta\Delta_\omega$ evaluated at $ev_FE/\omega=\Delta_{00}$; the gap change
produced by an actual drive is this value multiplied by
$(ev_FE/\omega\Delta_{00})^2$.  Red is enhancement ($\delta\Delta_\omega>0$),
blue is suppression, and the solid black contour is $\delta\Delta_\omega=0$.
Energies are in units of the clean zero-temperature gap $\Delta_{00}$ and
temperatures in units of the $T_c$ of each panel.  The dashed curve marks, at each temperature, the frequency equal to the
maximum gap, $\omega=\Delta_{\rm max}(T)=\sqrt2\,\Delta(T)$, with $\Delta(T)$
the self-consistent solution of the equilibrium gap equation. It lets the
position of the enhancement region be read against the gap scale, which
shrinks toward $T_c$, rather than against the fixed $\Delta_{00}$ of the axis.  Remaining
parameters: inelastic scattering rate $\gamma_{\rm in}/\Delta_{00}=0.0144$,
pairing strength $\lambda=0.2$, cutoff $\omega_D/\Delta_{00}=7.20$.  The
enhancement region is a single strip attached to $T_c$, opening at
$T/T_c=0.87$ and $0.89$ respectively.  Its upper edge starts inside the gap,
crosses $\Delta_{\rm max}(T)$ at $T/T_c=0.89$ and $0.92$, and reaches
$1.5\,\Delta_{\rm max}(T)=2.2\,\Delta(T)$ and $1.3\,\Delta_{\rm max}(T)=
1.8\,\Delta(T)$ at the top of the range.  The growth of $|\delta\Delta_\omega|$
toward $T_c$ reflects the divergence of the pair susceptibility
$\chi_{\rm SH}(T)$, which increases by a factor of $26$ over the plotted range.}
\label{Fig-contour}
\end{figure*}

Assembling, the $\itt$ component of the full Keldysh function gives the following expression:
\beg\label{g2Kitt}
\begin{aligned}
&\frac{1}{2i}\textrm{Tr}\big[\hat{\tau}_2\gh_2^K\big]=\frac{1}{2i}\textrm{Tr}\Big(\hat{\tau}_2\big[\gh_2^{R}-\gh_2^{A}\big]\Big)t_\eps
\\&+\frac{g^R_{\bn\eps}S_1+f^R_{\bn\eps}S_0}{\etaR+\etaA}
-\frac{i\Gel}{2}\left(\frac{C_{\bn\eps}^{RA}Y_3-\overline{B}_{\bn\eps}^{RA}Y_2}{\etaR+\etaA}\right).
\end{aligned}
\en
Having derived this expression, we can now evaluate the static correction to the order parameter, Eq. \eqref{Self4Elig2K}. Note that since the pre-factor in front of $Y_2$ contains even powers of the $d$-wave form-factor, upon inserting it into the self-consistency condition it is easy to see that it will not contribute to $\delta\Delta_\omega$.

\section{Results}
Collecting all $\delta\Delta_\omega$ contributions we arrive to the following equation for the static second order correction to the pairing amplitude:
\beg\label{Form4Eli}
\chi_{\rm SH}^{-1}(T)\,\delta\Delta_\omega=\left(\frac{ev_FE}{\omega}\right)^{\!2}\Rcal(\omega),
\en
where $\chi_{\rm SH}^{-1}(T)$ is the static inverse pair susceptibility defined by \eqref{Ddef} in Appendix \ref{app:kernel} and 
\beg\label{R-kernel}
\Rcal(\omega)=\Rcal_{\rm reg.}(\omega)+\Rcal_{\rm anom.}(\omega) 
\en
is the dimensionless photo-kernel given by the sum of regular ($RA$) and anomalous (Keldysh) parts, Eqs. (\ref{Rreg},\ref{Ranom}) in Appendix \ref{app:kernel}. 

In Fig. \ref{Fig-Rreg} we show the frequency dependence of the regular part of the photo kernel \ref{R-kernel}. Note that function $R_{\rm reg.}$ remains negative for a wide range of model parameters. This means that it accounts for the pair breaking effects only. This should not be surprising since all contributions to $R_{\rm reg.}$ involve the equilibrium quasiparticle distribution function. 
In contrast to $R_{\rm reg.}(\omega)$, the anomalous part of the photo kernel becomes positive for a range of frequencies below $\Delta_{\rm max}$, Fig. \ref{Fig-Ranom}. Note that function $\Rcal_{\rm anom}(\omega)$, the response of the nonequilibrium
distribution, is zero in a clean system but not for the reason the explicit factors suggest. Indeed, only the
second line of \eref{Ranom} carries $1/\tau$ in front. The first line, which
is the direct drive contribution to the nonequilibrium distribution, has no
explicit factor of the elastic rate, and its denominator
$\etaR+\etaA=2i\,\Im\,\etaR$ stays of order $\gin$ as $\Gel\to0$. That line
vanishes instead through a cancellation between absorption and emission processes, so that the first term becomes proportional to $\Gel$ as well. 
Enhancement is therefore not a competition
between two comparable terms but a single condition, $\Rcal_{\rm
anom}>|\Rcal_{\rm reg}|\simeq1$, and it requires disorder for the same reason
the $s$-wave effect does: at $\bk=0$ a spatially uniform drive on a clean
superconductor produces no absorption.  

Since the thermodynamic stability requires that $\chi_{\rm SH}^{-1}>0$, the sign of $\delta\Delta_\omega$ is determined by the sign of the photo-kernel. At temperatures close to $T_c$ the photo-kernel can be expanded in powers of
the order parameter at fixed drive frequency,
\beg\label{PhotoKernelTc}
{\cal R}(\omega)=A(\omega)\Delta+B(\omega)\Delta^3+O(\Delta^5),
\en
so that the sign of $\delta\Delta_\omega$ at leading order in $\Delta$ is the
sign of $A(\omega)$.  In the clean limit $A(\omega)$ vanishes identically
\cite{Steineman2026}.  At finite $\Gel$ it does not: we find $A(\omega)<0$ at
every frequency and every disorder strength we have examined, so that at fixed
$\omega$ the drive suppresses the order parameter sufficiently close to $T_c$.
This is not in conflict with the enhancement seen in Fig.~\ref{Fig-contour}
up to $T/T_c=0.98$.  The expansion \eqref{PhotoKernelTc} holds for
$\Delta(T)\ll\omega$, whereas the enhancement region lies at $\omega\simeq
(0.7$--$2.2)\,\Delta(T)$ and moves to lower frequency with the gap.  For any
fixed $\omega$ the upper edge of the region therefore passes below $\omega$
as $T\to T_c$, and $\Rcal(\omega)$ changes sign as $A(\omega)<0$ requires.
Along a frequency that tracks the gap, $\omega\propto\Delta(T)$, the region
persists much closer to $T_c$; it must still close, because the ratios
$\gin/\Delta(T)$ and $\Gel/\Delta(T)$ that bound it from above grow without
limit as $\Delta\to0$, but this occurs within a fraction of a percent of
$T_c$, above the range shown.  The near-$T_c$ behavior at fixed $\omega$ is
thus governed by $A(\omega)$, while the structure of Fig.~\ref{Fig-contour}
is governed by $B(\omega)$ and higher terms, and the two descriptions refer
to different corners of the $(\omega,T)$ plane.

We have also added the dashed curve in Fig.~\ref{Fig-contour} which serves as a frequency reference. The
axes of the figure are fixed in units of $\Delta_{00}$, while the natural
scale for the drive frequency is the gap $\Delta(T)$,
which shrinks toward $T_c$. At each temperature the dashed curve marks the
frequency equal to the maximum gap, $\omega=\Delta_{\rm max}\equiv\sqrt2\,\Delta(T)$,
so that the frequency and temperature position of the enhancement region can be read against the gap
scale directly. The choice of $\Delta_{\max}$ is made for the comparison
with the $s$-wave case, where the enhancement region is bounded above by
the pair-breaking threshold $\omega=2\Delta$ set by the $s$-wave spectrum
\cite{Eliashberg1970,Ivlev1971}: with frequency below the edge photons cannot break pairs and the
redistribution of quasiparticles raises the gap, while for $\omega$ above photons induce pair breaking thus suppressing
the paring. In a $d$-wave superconductor there is obviously no such threshold due to the nodal quasiparticles, so the only spectral feature that survives is the maximum gap at the antinodes, and the dashed curve asks whether the window respects it: as we can see from our results it does not. Both edges of the enhancement region scale with $\Delta(T)$ ---
the region moves to lower frequency as the gap collapses, which is why it
leans toward the temperature axis in the figure --- but neither edge is
pinned to $\Delta_{\max}$. Moreover, the lower edge, which has no counterpart in the
$s$-wave case, is present at every temperature: it lies at
$\omega\simeq\Delta(T)$ near the onset and falls to $0.2\,\Delta(T)$ and
$0.5\,\Delta(T)$ at $T/T_c=0.98$, because nodal quasiparticles absorb at
arbitrarily small $\omega$ and heat the quasiparticle system, so that there
is no adiabatic corner in which the drive is harmless. So the important physics take away for us is that the edges of the
$d$-wave window are thus set by the competition between redistribution,
heating and pair breaking in the kinetics, not by a feature of the spectrum.

\begin{table}[b]
\caption{Largest attainable $\delta T_c$ at fixed drive amplitude, and the
temperature and frequency at which it is reached, for $\Gel/\Delta_{00}=0.072$;
$\delta T_c$ is the temperature shift equivalent to the computed gap shift,
$\delta T_c=\delta\Delta_\omega/|d\Delta/dT|$, with $\Delta(T)$ the
self-consistent gap.  The entries are restricted to
$\delta\Delta_\omega\lesssim0.1\,\Delta$, which guarantees the applicability
of the perturbation theory.}
\label{tab:best}
\begin{ruledtabular}
\begin{tabular}{cccc}
$ev_FE/\omega\Delta_{00}$ & $\delta T_c/T_c$ & $T/T_c$ & $\omega/\Delta_{00}$ \\ \hline
0.05 & 0.0050 & 0.972 & 0.282 \\
0.10 & 0.0103 & 0.950 & 0.282 \\
0.15 & 0.0145 & 0.931 & 0.506 \\
0.20 & 0.0178 & 0.916 & 0.529 \\
\end{tabular}
\end{ruledtabular}
\end{table}

The quantity shown in Fig.~\ref{Fig-contour} is $\delta\Delta_\omega$ at
$(ev_FE/\omega)^2=1$.  For an actual drive the gap shift is obtained by
multiplying it by $(ev_FE/\omega)^2$, which, with $\hbar=1$, has units of an energy
squared and is measured here in units of $\Delta_{00}^2$:
\beg\label{drive-energy}
\frac{ev_FE}{\omega}\equiv\alpha\,\Delta_{00},\qquad
\frac{\delta\Delta_\omega}{\Delta(T)}=s(T,\omega)\,\alpha^2 ,
\en
where $s$ is the value read off the figure divided by $\Delta(T)/\Delta_{00}$.
Because the linearized theory requires $\delta\Delta_\omega\ll\Delta(T)$, the
drive strength $\alpha$ for which it applies is small.  The large values on the
color scale of Fig.~\ref{Fig-contour} near $T_c$ are therefore values of the
coefficient $s(T,\omega)$ in Eq.~\eqref{drive-energy}, obtained by setting
$\alpha=1$. The gap change produced by an admissible drive is $s\,\alpha^2$ times
$\Delta(T)$, and is thus smaller than $\Delta(T)$ by construction. 

From our results it also follows that the possible maximum value of $\delta T_c$ saturates. Indeed, increasing the magnitude of the drive 
pushes the temperature at which the linear treatment remains
valid further below $T_c$, where the inverse pair susceptibility
$\chi_{\rm SH}^{-1}$ is larger and the same kernel produces a smaller shift. As a result $\delta T_c/T_c$ grows
only from $0.5\%$ to about $1.8\%$ as $ev_FE/\omega$ goes from $0.05\Delta_{00}$ to $0.20\Delta_{00}$.
The optimum sits at $T$ between $0.92\,T_c$ and $0.97\,T_c$ and at drive
frequencies $\omega\simeq0.3$--$0.5\,\Delta_{00}$.  The effect is therefore a percent-level shift
of $T_c$, and it is nearly independent of disorder within the window.

\section{Discussion}
The constraint $\delta\Delta_\omega/\Delta\le0.1$ that defines the entries of
Table~\ref{tab:best} is what makes them finite.  This constraint is needed here since without it
$\delta\Delta_\omega\propto\chi_{\rm SH}$ grows without bound toward $T_c$, because the inverse
pair susceptibility $\chi_{\rm SH}^{-1}(T)$ that multiplies it in
Eq.~\eqref{Form4Eli} goes to zero there.  That growth is real, but it is not
captured by a calculation carried to first order in the intensity $O(E^2)$ of the
drive, since the enhanced gap then feeds back on the distribution that
produced it.  Within the window the result is nearly independent of
disorder: at $\Gel/\Delta_{00}=0.252$ the same construction gives
$\delta T_c\simeq0.010\,T_c$ at $T=0.95\,T_c$ for $ev_FE/\omega=0.1\,\Delta_{00}$,
against $0.011\,T_c$ at $\Gel/\Delta_{00}=0.072$.

We now estimate the electric field to which these numbers correspond. Since
$ev_FE/\omega=\alpha\,\Delta_{00}$, the field amplitude is
\beg\label{E0}
E=\alpha\,\frac{\omega E_0}{\Delta_{00}},\qquad
E_0\equiv\frac{\Delta_{00}^2}{\hbar e v_F},
\en
where $E_0$ is the field at which the drive energy $ev_FE/\omega$ equals
$\Delta_{00}$ for $\omega=\Delta_{00}$. For cuprate-like parameters,
$T_c=90$~K with the gap ratio of the present model,
$\Delta_{00}=1.576\,T_c=12.2$~meV, and $\hbar v_F=1.5$~eV\,\AA, this gives
$E_0\simeq10$~kV/cm, and the $1\%$ shift of $T_c$ in the second row of
Table~\ref{tab:best} requires $E\simeq0.28$~kV/cm at $\simeq0.8$~THz.
Because $E_0\propto\Delta_{00}^2/v_F$, the corresponding field for a
conventional low-$T_c$ film is far smaller: for Al, with $T_c=1.2$~K and
$\hbar v_F=13.6$~eV\,\AA, $E_0\simeq20$~V/m and the required amplitude is
below $1$~V/m at about $10$~GHz, the regime in which microwave stimulation
was originally observed \cite{Dayem1967,Wyatt1966}. Two caveats attach to
these numbers. The conversion uses the gap ratio of the present model; with
the measured antinodal gap of an optimally doped cuprate, roughly twice
$\sqrt2\,\Delta_{00}$, the field scale rises as $\Delta_{00}^2$, by a factor
of four to five. And the amplitudes are one to three orders of magnitude
below the pump fields typical of light-induced superconductivity
experiments, which therefore lie well outside the perturbative regime used
here.
 
Three limitations of the formalism provide constraints on the validity of these statements. First, the
response is computed to second order in the field, that is to first order in
the intensity $(ev_FE/\omega)^2$, so no feedback of the enhanced gap on the
quasiparticle distribution has been included. Second, inelastic relaxation is the
reservoir model of \eref{sigma-in}, in which $\gin$ is a single
phenomenological rate; since the effect scales as $1/\gin$, a microscopic
collision integral would change the magnitude of the window even where it
does not move its edges. Third, scattering on potential impurities is treated
in the Born limit, whereas scattering in the cuprates is often closer to
unitary. In this regard the heavy-fermion superconductors of the Ce-115
family, in which the elastic rate can be tuned by chemical substitution, may
be a more suitable testing platform \cite{SarraoThompsonReview}.

\section{Conclusions}
In this work we have shown that a disordered $d$-wave superconductor does support
photo-enhancement of the order parameter, and that the region in which it does
is bounded in every physical parameter of the problem rather than being the simple
threshold $\omega<2\Delta$ of the $s$-wave case. The enhancement region represented by a single connected strip has been computed in the $(\omega,T)$ plane using the self-consistent $\Delta(T)$. We have demonstrated that the enhancement is carried entirely by the anomalous kernel $\Rcal_{\rm
anom.}(\omega)$ which is determined by the deviations of the quasiparticle distribution function from equilibrium. The regular part of the photo-kernel,
$\Rcal_{\rm reg.}(\omega)$, which describes the spectral response of the driven propagators, is negative at every
frequency, temperature and disorder strength we have examined, and it is
nearly featureless. For example, at temperatures $T\simeq\Delta$ it remains of order $-1$ across the
whole frequency range.

Our theoretical framework is a perturbative analysis of the Eilenberger
equation for a disordered $d$-wave superconductor, carried to second order in
the external electromagnetic field for the retarded, advanced and Keldysh propagators, with inelastic
relaxation and elastic scattering included and the full second-order impurity
vertex retained. We emphasize that for the Eliashberg effect the impurity vertex is not
optional: it carries the conservation law that protects the isotropic
component of the nonequilibrium distribution and, as we have shown above,
dropping it replaces the $1/\gin$ relaxation of that component by $1/\tau$.
The $d$-wave problem is kept tractable by one exact property: the anomalous
propagator averages to zero over the Fermi surface, so impurities renormalize
the quasiparticle energy but not the pairing field, and the disordered ground
state is fixed by a single scalar equation.

We believe it is worth emphasizing that with regard to the Eliashberg effect
a nodal superconductor is special in that the disorder which opens the
channel for the enhancement also breaks pairs, so that potential scattering
becomes a double-edged sword: too little disorder leaves the quasiparticle
redistribution channel closed and too much destroys the gap it would enhance.
In the case that we have considered $\Rcal(\omega)$ is positive only between
$\Gel\simeq0.011\Delta_{00}$ and $\Gel\simeq0.40\Delta_{00}$ and is largest
near $\Gel\simeq0.14\Delta_{00}$ --- the value the two panels of
Fig.~\ref{Fig-contour} bracket. Nevertheless, nothing in our formalism
is tied to a $d$-wave form factor, since $\Ycal_\bn$ enters only through the
angular averages.

As for the experimental consequences, we believe that our theory provides both a constraint on where to look for the effect and a
prediction of its magnitude. The window occupies the upper $10$--$13\%$ of the
temperature range and sits at frequencies of order $\Delta(T)$, a scale that
itself collapses as $T\to T_c$, so a fixed-frequency drive does not stay
inside it. Within the window the shift is at the percent level, $\delta T_c\simeq0.01\,T_c$
for $ev_FE/\omega=0.1\,\Delta_{00}$, which for cuprate parameters is a field of
a few hundred V/cm at about $1$~THz (Table~\ref{tab:best}).
Therefore, the frequency of the external radiation must be tuned to a temperature-dependent pairing gap, and a
temperature sweep at fixed $\omega$ crosses the window rather than tracking
it. The disorder requirement is equally well defined, and a sample either too clean
or too dirty shows suppression at every frequency. Taken together these conditions set a concrete target for experiments on
light-induced superconductivity in the cuprates: the mechanism is available in
a nodal superconductor --- unlike in the clean limit, where it vanishes at all
frequencies \cite{Steineman2026} --- but only inside a corner of parameter space
that a measurement has to be designed to reach.

Several avenues for future exploration follow directly from our work. The first is to explore the rest of the
$(\omega,T,\Gel)$ phase space: two elastic rates have been mapped on the full
plane, and the two-sided structure in $\Gel$ was located at fixed $\Delta$, so
the edges of the dome on the self-consistent trajectory have not been drawn yet. The
second avenue consists in venturing beyond perturbative second-order response and beyond the escape model for $\gin$ --- a
genuine collision integral for $\gin$, and the nonequilibrium steady state at
finite drive, in which the enhanced gap feeds back on the distribution that
produced it --- together with unitary rather than Born scatterers. The third
is the extension to finite photon momentum and to the Eliashberg superconductors \cite{MyPrecious} as well as to other order-parameter
symmetries such as the $p$-wave states relevant near ferromagnetic quantum critical points \cite{Dzero2004}, for which the construction carries over directly in the unitary equal-spin case and otherwise requires the spin structure of the triplet gap to be restored.

\begin{acknowledgments}
This work was financially supported by the National Science
Foundation Grant No. DMR-2400484. MD thanks the Niels Bohr Institute,
where a part of this work was completed, for its hospitality. 
Most of the numerical calculations were performed using an allocation of computing time and other resources of the Ohio Supercomputer Center~\cite{OSC1987}. The authors acknowledge the use of Anthropic's
Claude for cross-checking the analytical derivations, for independent
numerical verification of the results, and for figure preparation and
manuscript checking. All outputs of Claude were verified and validated by
the authors.
\end{acknowledgments}

\section*{Data availability} The data that support the findings of this article were generated by numerical evaluation of the analytical expressions derived herein and can be regenerated in full from the information given in the paper. For this reason the authors did not  deposit them in a public repository. The data and the codes that produce figures presented in this manuscript are available from the authors upon request. 

%
%
\begin{appendix}
\begin{widetext}

\section{First-order corrections}\label{app:first}
For the $+$ harmonic the first-order equation at $\bk=0$ is
\beg\label{A-Eq4g1}
\big[\eps\ttau3+\hat\Delta_\bn,\gh_{1,+}\big]+\frac\omega2\big\{\ttau3,\gh_{1,+}\big\}
+\frac{i\gin}{2}\big[\ttau3,\gh_{1,+}\big]+\frac{i\Gel}{2}\big[\Gcal_\eps\ttau3,\gh_{1,+}\big]_\circ
={\cal A}_+\Big[\gh_{\bn\eps+\frac\omega2}\ttau3-\ttau3\gh_{\bn\eps-\frac\omega2}\Big],
\en
with ${\cal A}_\pm$ of \eref{amps}, the impurity commutator being
$[\avg{\gh_0},\gh_1]$ only, since $\avg{\gh_1}=0$. By the shift rule stated below
\eref{FSaverage} the left-hand side equals
$\tveps_{\eps+\frac\omega2}\ttau3\gh_{1,+}-\gh_{1,+}\ttau3\tveps_{\eps-\frac\omega2}+[\hat\Delta_\bn,\gh_{1,+}]$,
i.e.\ both self-energies are absorbed into $\tveps$ of \eref{xitilde} at the two
shifted energies. Using $\tveps_a\ttau3+\hat\Delta_\bn=\eta_a\gh_a$ this is
$\eta_{a}\gh_a\gh_{1,+}-\eta_b\gh_{1,+}\gh_b$ with $a=\eps+\frac\omega2$,
$b=\eps-\frac\omega2$, and together with the first-order normalization
$\gh_a\gh_{1,+}+\gh_{1,+}\gh_b=0$ following from \eref{Normg} it becomes
$(\eta_a+\eta_b)\gh_a\gh_{1,+}$. Multiplying by $\gh_a$ gives \eref{g1RA}; the $-$
harmonic follows from $\omega\to-\omega$, $\bE\to\bE^*$. For the Keldysh component
one inserts the ansatz \eref{g1K} into the Keldysh block of \eref{A-Eq4g1}; the
terms proportional to $t_{\eps\pm\omega/2}$ cancel against the $R$ and $A$
equations once the bath self-energies \eref{sigma-in} and the $\hs^K$ of the
impurities, $-i\Gel\Re\Gcal_\eps\,t_\eps\ttau3$, are included, and what remains is
solved as before with $\eta^R_a+\eta^A_b$ in the denominator, giving
$\delta\gh^K_{1,\pm}$ of \eref{g1K}. Both closed forms agree with a direct
numerical solution of the equations of motion plus normalization to $10^{-15}$.

The block identity \eref{SimpleBCdef} follows from
$\gh_a\ttau3\gh_b=(g_ag_b+f_af_b)\ttau3+(g_af_b+f_ag_b)\itt$, which uses
$\ttau3\,\itt=\ttau1=-\itt\,\ttau3$ and $(\itt)^2=-1$. In the notation of
\eref{BCdef},
\begin{align}
\gh_{1,\pm}^{R(A)}(\bn\eps)&=-{\cal A}_\pm\frac{B^{R(A)}_{\bn}(\eps+\frac\omega2,\eps-\frac\omega2)\ttau3
+C^{R(A)}_{\bn}(\eps+\frac\omega2,\eps-\frac\omega2)\itt}{\eta^{R(A)}_{\bn\eps+\frac\omega2}+\eta^{R(A)}_{\bn\eps-\frac\omega2}},\label{A-blocks}\\
\delta\gh^K_{1,\pm}(\bn\eps)&=-{\cal A}_\pm\frac{\big[B^{RA}_\bn(\eps\pm\frac\omega2,\eps\mp\frac\omega2)\ttau3
+C^{RA}_\bn(\eps\pm\frac\omega2,\eps\mp\frac\omega2)\itt\big]\big(t_{\eps\pm\frac\omega2}-t_{\eps\mp\frac\omega2}\big)}
{\eta^{R}_{\bn\eps\pm\frac\omega2}+\eta^{A}_{\bn\eps\mp\frac\omega2}},\label{A-blocksK}
\end{align}
where $B^{RA}_\bn(a,b)=g^R_ag^A_b+f^R_af^A_b-1$ and
$C^{RA}_\bn(a,b)=g^R_af^A_b+f^R_ag^A_b$ (first argument retarded, second
advanced), while the single-branch $B$, $C$ of \eref{BCdef} are symmetric in their
arguments.

Throughout the appendices we write $[\hat X]_{\ttau3}$ and $[\hat X]_{\itt}$ for the
coefficients of $\ttau3$ and $\itt$ in the expansion of a Nambu matrix over
$\{\ttau0,\ttau1,\itt,\ttau3\}$; in terms of the traces used in the main text,
$[\hat X]_{\ttau3}=\frac12\Tr[\ttau3\hat X]$ and
$[\hat X]_{\itt}=\frac{1}{2i}\Tr[\hat\tau_2\hat X]=-\frac12\Tr[\itt\,\hat X]$.

\section{Second-order spectral functions}\label{app:RA}
\paragraph{Source terms.}
The drive source and the normalization term in \eref{eq:g2RA} are
\begin{align}
\hat{\cal R}&={\cal A}_+\Big[\hat g^{\,-}_{1}(\eps+\tfrac\omega2)\ttau3-\ttau3\,\hat g^{\,-}_{1}(\eps-\tfrac\omega2)\Big]
+{\cal A}_-\Big[\hat g^{\,+}_{1}(\eps-\tfrac\omega2)\ttau3-\ttau3\,\hat g^{\,+}_{1}(\eps+\tfrac\omega2)\Big],\label{B-R}\\
\hat{\cal Z}&=\hat g^{\,+}_1(\eps-\tfrac\omega2)\,\hat g^{\,-}_1(\eps-\tfrac\omega2)
+\hat g^{\,-}_1(\eps+\tfrac\omega2)\,\hat g^{\,+}_1(\eps+\tfrac\omega2),\label{B-Z}
\end{align}
where $\hat g^{\,\pm}_1\equiv\gh_{1,\pm}/{\cal A}_\pm$ are the amplitude-stripped
blocks of \eref{A-blocks}, and $\hat{\cal Z}$ is the static part of
$(\gh_1\circ\gh_1)$ that appears in the second-order expansion of \eref{Normg}.
Each block in (\ref{B-R})--(\ref{B-Z}) involves the pair of energies
$(\eps,\eps\pm\omega)$: $\hat g_1^{\,-}(\eps+\frac\omega2)$ and
$\hat g_1^{\,+}(\eps+\frac\omega2)$ both carry $(\eps,\eps+\omega)$ and are equal on
a single branch, and likewise
$\hat g_1^{\,-}(\eps-\frac\omega2)=\hat g_1^{\,+}(\eps-\frac\omega2)$ carry
$(\eps,\eps-\omega)$. Writing $B_s\equiv B_\bn(\eps,\eps+\sw)$,
$C_s\equiv C_\bn(\eps,\eps+\sw)$ and $\eta_s\equiv\eta_{\bn\eps+\sw}$, the
commutator $[-(B\ttau3+C\itt),\ttau3]=2C\ttau1$ and the square
$(B\ttau3+C\itt)^2=B^2-C^2$ give
\beg\label{B-RZ}
\hat{\cal R}=2{\cal A}_+{\cal A}_-\,\ttau1\sum_{s=\pm}\frac{C_s}{\eta_{\bn\eps}+\eta_s},\qquad
\hat{\cal Z}={\cal A}_+{\cal A}_-\,\ttau0\sum_{s=\pm}\frac{B_s^2-C_s^2}{(\eta_{\bn\eps}+\eta_s)^2}
=-2{\cal A}_+{\cal A}_-\,\ttau0\sum_{s=\pm}\frac{B_s}{(\eta_{\bn\eps}+\eta_s)^2},
\en
the last step using the identity $B^2-C^2=-2B$, which follows from
$(g_ag_b+f_af_b)^2-(g_af_b+f_ag_b)^2=(g_a^2-f_a^2)(g_b^2-f_b^2)=1$.

\paragraph{Vertex response and drive components.}
The particular-solution formula quoted below \eref{g2RAsol} acts on the four Nambu
generators as
\beg\label{B-resp}
{\cal V}[\ttau0]=0,\qquad
{\cal V}[\ttau1]=-\frac{\ttau1}{\eta_{\bn\eps}},\qquad
{\cal V}[\ttau3]=\frac{f_{\bn\eps}^2}{\eta_{\bn\eps}}\ttau3+\frac{g_{\bn\eps}f_{\bn\eps}}{\eta_{\bn\eps}}\itt,\qquad
{\cal V}[\itt]=-\frac{1+f_{\bn\eps}^2}{\eta_{\bn\eps}}\itt-\frac{g_{\bn\eps}f_{\bn\eps}}{\eta_{\bn\eps}}\ttau3,
\en
with ${\cal V}[\hat G]=(\gh_{\bn\eps}\hat G\gh_{\bn\eps}-\hat G)/2\eta_{\bn\eps}$ as
defined above \eref{X3closure}; the first two follow from \eref{Normg} and
$g^2-f^2=1$, the last two from
$\gh\ttau3\gh=(g^2+f^2)\ttau3+2gf\,\itt$ and
$\gh\,\itt\,\gh=-(g^2+f^2)\itt-2gf\,\ttau3$ together with
$g^2+f^2-1=2f^2$ and $g^2+f^2+1=2(1+f^2)=2g^2$.

The drive part of \eref{g2RAsol} is
\beg\label{B-g2}
\frac{\gh_{\bn\eps}(\hat{\cal R}-\eta_{\bn\eps}\hat{\cal Z})}{2\eta_{\bn\eps}}
={\cal A}_+{\cal A}_-\sum_{s=\pm}\left[\frac{C_s}{\eta_{\bn\eps}(\eta_{\bn\eps}+\eta_s)}\,\gh_{\bn\eps}\ttau1
+\frac{B_s}{(\eta_{\bn\eps}+\eta_s)^2}\,\gh_{\bn\eps}\right],
\qquad \gh_{\bn\eps}\ttau1=f_{\bn\eps}\ttau3+g_{\bn\eps}\itt ,
\en
manifestly in the span of $\{\ttau3,\itt\}$. Adding the $\dDh_\bn$ term of
\eref{g2RAsol}, which by \eref{B-resp} with
$\dDh_\bn=\itt\,\Ycal_\bn\delta\Delta_\omega$ entering with the sign opposite to
$\hs_2$ contributes
$(\dDh_\bn-\gh\dDh_\bn\gh)/2\eta=\Ycal_\bn\delta\Delta_\omega(g^2\,\itt+gf\,\ttau3)/\eta$,
gives the two components that source the closures \eref{X3closure},
\begin{align}
\big[\gh_2^{(0)}\big]_{\itt}&=\delta\Delta_\omega\,\Ycal_\bn\,\frac{g_{\bn\eps}^2}{\eta_{\bn\eps}}
+\ampl\sum_{s=\pm}\left[\frac{g_{\bn\eps}\,C_{s}}{\eta_{\bn\eps}(\eta_{\bn\eps}+\eta_{s})}
+\frac{f_{\bn\eps}\,B_{s}}{(\eta_{\bn\eps}+\eta_{s})^{2}}\right],\label{g20itt}\\
\big[\gh_2^{(0)}\big]_{\ttau3}&=\delta\Delta_\omega\,\Ycal_\bn\,\frac{g_{\bn\eps}f_{\bn\eps}}{\eta_{\bn\eps}}
+\ampl\sum_{s=\pm}\left[\frac{f_{\bn\eps}\,C_{s}}{\eta_{\bn\eps}(\eta_{\bn\eps}+\eta_{s})}
+\frac{g_{\bn\eps}\,B_{s}}{(\eta_{\bn\eps}+\eta_{s})^{2}}\right],\label{g20t3}
\end{align}
identically for $R$ and $A$, and with no $\ttau0$ or $\ttau1$ component. The two
expressions differ only by the exchange $g_{\bn\eps}\leftrightarrow f_{\bn\eps}$ of
the single-energy prefactors, which is a useful implementation cross-check.

That \eref{B-g2} solves \eref{eq:g2RA} is verified directly: with
$\tveps\ttau3+\hat\Delta_\bn=\eta\gh$, the commutator of the first term is
$\eta[\gh,\gh\ttau1]C_s/[\eta(\eta+\eta_s)]=(\ttau1-\gh\ttau1\gh)C_s/(\eta+\eta_s)=2\ttau1C_s/(\eta+\eta_s)$,
using $\gh\ttau1\gh=-\ttau1$, while the second term commutes with $\gh$; and
\eref{Norm4g2RA} is satisfied because $\gh(\gh\ttau1)+(\gh\ttau1)\gh=0$ and
$2\gh\gh\,B_s/(\eta+\eta_s)^2=2B_s/(\eta+\eta_s)^2$.

\paragraph{Closure.}
The vertex part of \eref{g2RAsol} is \eref{B-resp} with the amplitudes
$-\frac{i\Gel}{2}X_a$; its components are
\beg\label{B-g21}
\big[\gh_2^{(1)}\big]_{\ttau1}=\frac{i\Gel}{2}\frac{X_1}{\eta_{\bn\eps}},\qquad
\big[\gh_2^{(1)}\big]_{\itt}=\frac{i\Gel}{2}X_2\frac{1+f_{\bn\eps}^2}{\eta_{\bn\eps}}-\frac{i\Gel}{2}X_3\frac{g_{\bn\eps}f_{\bn\eps}}{\eta_{\bn\eps}},\qquad
\big[\gh_2^{(1)}\big]_{\ttau3}=-\frac{i\Gel}{2}X_3\frac{f_{\bn\eps}^2}{\eta_{\bn\eps}}+\frac{i\Gel}{2}X_2\frac{g_{\bn\eps}f_{\bn\eps}}{\eta_{\bn\eps}},
\en
and $[\gh_2^{(1)}]_{\ttau0}=0$. Averaging $\gh_2=\gh_2^{(0)}+\gh_2^{(1)}$ over the
Fermi surface and equating components gives
$X_0=\avg{[\gh_2^{(0)}]_{\ttau0}}=0$,
$X_1\big[1-\frac{i\Gel}{2}\avg{\eta_{\bn\eps}^{-1}}\big]=\avg{[\gh_2^{(0)}]_{\ttau1}}=0$,
and the pair \eref{X3closure}, the $gf/\eta$ cross terms dropping because they are
odd in $\Ycal_\bn$ while $g$, $f^2$ and $\eta$ are even. Note that the
$\delta\Delta_\omega$ term of \eref{g20t3} does \emph{not} drop from the $X_3$
closure, since it is itself proportional to $\Ycal_\bn$ and
$\avg{\Ycal_\bn\,g f/\eta}\neq0$; this is the origin of the impurity-vertex
correction to the pair susceptibility in Appendix~\ref{app:kernel}.

\section{Second-order Keldysh function}\label{app:K}
\paragraph{Drive part.}
Inserting the ansatz \eref{g2Kansatz} into the Keldysh block of the second-order
equation and into the Keldysh normalization condition, and using the first-order
Keldysh functions \eref{g1K}, all terms proportional to $t_\eps$ times the $R$ and
$A$ equations cancel, and $\delta\gh_2^{K,0}$ obeys
\beg\label{C-eq}
\tveps^R_\eps\ttau3\,\delta\gh_2^{K,0}-\delta\gh_2^{K,0}\ttau3\,\tveps^A_\eps+\big[\hat\Delta_\bn,\delta\gh_2^{K,0}\big]
=\hat{\cal R}^K+\hat{\cal R}^{\delta K},\qquad
\gh^R_{\bn\eps}\delta\gh_2^{K,0}+\delta\gh_2^{K,0}\gh^A_{\bn\eps}=-\hat{\cal D},
\en
with (all blocks amplitude-stripped, so that the right-hand sides carry
${\cal A}_+{\cal A}_-$)
\begin{align}
\hat{\cal R}^K&=\Big[\hat g_1^{\,-R}(\eps+\tfrac\omega2)\ttau3\,(t_{\eps+\omega}-t_\eps)
-\ttau3\,\hat g_1^{\,-A}(\eps-\tfrac\omega2)(t_\eps-t_{\eps-\omega})\Big]\nonumber\\
&\quad+\Big[\hat g_1^{\,+R}(\eps-\tfrac\omega2)\ttau3\,(t_{\eps-\omega}-t_\eps)
-\ttau3\,\hat g_1^{\,+A}(\eps+\tfrac\omega2)(t_\eps-t_{\eps+\omega})\Big],\label{C-RK}\\
\hat{\cal R}^{\delta K}&=\Big[\delta\hat g_1^{\,-}(\eps+\tfrac\omega2)\ttau3-\ttau3\,\delta\hat g_1^{\,-}(\eps-\tfrac\omega2)\Big]
+\Big[\delta\hat g_1^{\,+}(\eps-\tfrac\omega2)\ttau3-\ttau3\,\delta\hat g_1^{\,+}(\eps+\tfrac\omega2)\Big],\label{C-RdK}\\
\hat{\cal D}&=\hat g_1^{\,+R}(\eps-\tfrac\omega2)\,\delta\hat g_1^{\,-}(\eps-\tfrac\omega2)
+\hat g_1^{\,-R}(\eps+\tfrac\omega2)\,\delta\hat g_1^{\,+}(\eps+\tfrac\omega2)\nonumber\\
&\quad+\delta\hat g_1^{\,+}(\eps-\tfrac\omega2)\,\hat g_1^{\,-A}(\eps-\tfrac\omega2)
+\delta\hat g_1^{\,-}(\eps+\tfrac\omega2)\,\hat g_1^{\,+A}(\eps+\tfrac\omega2).\label{C-D}
\end{align}
The solution is \eref{dg2K0}. Since $\tveps^{R}\ttau3+\hat\Delta_\bn=\etaR\gh^R$ and
$\tveps^{A}\ttau3+\hat\Delta_\bn=\etaA\gh^A$, the left-hand side of \eref{C-eq}
acting on $\delta\gh_2^{K,0}=\gh^R\hat{\cal S}/(\etaR+\etaA)$ is
$(\etaR\hat{\cal S}-\etaA\gh^R\hat{\cal S}\gh^A)/(\etaR+\etaA)$, while the
normalization condition reads
$(\hat{\cal S}+\gh^R\hat{\cal S}\gh^A)/(\etaR+\etaA)=-\hat{\cal D}$. Eliminating
$\gh^R\hat{\cal S}\gh^A$ between the two turns the left-hand side into
$\hat{\cal S}+\etaA\hat{\cal D}$, whence
$\hat{\cal S}=\hat{\cal R}^K+\hat{\cal R}^{\delta K}-\etaA\hat{\cal D}$, as quoted in
\eref{dg2K0}. 

Every block in (\ref{C-RK})--(\ref{C-D}) is of the form $B\ttau3+C\itt$, and
$(B\ttau3+C\itt)\ttau3=B-C\ttau1$, $\ttau3(B\ttau3+C\itt)=B+C\ttau1$,
$(B_1\ttau3+C_1\itt)(B_2\ttau3+C_2\itt)=(B_1B_2-C_1C_2)+(B_1C_2-C_1B_2)\ttau1$,
so $\hat{\cal S}=S_0\ttau0+S_1\ttau1$ exactly, as asserted below \eref{dg2K0}. With
the abbreviations
\beg\label{C-abbrev}
\begin{aligned}
&B^R_s=g^Rg^R_s+f^Rf^R_s-1, && C^R_s=g^Rf^R_s+f^Rg^R_s, &&
B^A_s=g^Ag^A_s+f^Af^A_s-1, && C^A_s=g^Af^A_s+f^Ag^A_s,\\
&\tilde B_s=g^Rg^A_s+f^Rf^A_s-1, && \tilde C_s=g^Rf^A_s+f^Rg^A_s, &&
B^K_s=g^Ag^R_s+f^Af^R_s-1, && C^K_s=g^Af^R_s+f^Ag^R_s,
\end{aligned}
\en
in which unsubscripted functions are at $(\bn\eps)$, the subscript $s$ denotes the
argument $\eps+\sw$, and $\delta t_s\equiv t_\eps-t_{\eps+\sw}$, the two scalars are
(at unit amplitude, ${\cal A}_+{\cal A}_-\to1$)
\begin{align}
S_0&=\sum_{s=\pm}\delta t_s\left\{\frac{B^R_s}{\etaR+\eta^R_s}+\frac{B^A_s}{\etaA+\eta^A_s}
-\frac{\tilde B_s}{\etaR+\eta^A_s}-\frac{B^K_s}{\etaA+\eta^R_s}\right.\nonumber\\
&\qquad\qquad\left.+\etaA\left[\frac{B^R_sB^K_s-C^R_sC^K_s}{(\etaR+\eta^R_s)(\etaA+\eta^R_s)}
-\frac{\tilde B_sB^A_s-\tilde C_sC^A_s}{(\etaR+\eta^A_s)(\etaA+\eta^A_s)}\right]\right\},\label{S0}\\
S_1&=\sum_{s=\pm}\delta t_s\left\{-\frac{C^R_s}{\etaR+\eta^R_s}+\frac{C^A_s}{\etaA+\eta^A_s}
+\frac{\tilde C_s}{\etaR+\eta^A_s}-\frac{C^K_s}{\etaA+\eta^R_s}\right.\nonumber\\
&\qquad\qquad\left.+\etaA\left[\frac{B^R_sC^K_s-C^R_sB^K_s}{(\etaR+\eta^R_s)(\etaA+\eta^R_s)}
-\frac{\tilde B_sC^A_s-\tilde C_sB^A_s}{(\etaR+\eta^A_s)(\etaA+\eta^A_s)}\right]\right\}.\label{S1}
\end{align}
The four single terms in each brace come from $\hat{\cal R}^K$ (first two) and
$\hat{\cal R}^{\delta K}$ (last two); the bracket comes from $-\etaA\hat{\cal D}$.
Equations \eref{dg2K0comp} then follow from $\gh^R\ttau0=g^R\ttau3+f^R\itt$ and
$\gh^R\ttau1=f^R\ttau3+g^R\itt$.

\paragraph{Explicit form of the Keldysh drive components.}
For implementation it is convenient to eliminate $S_0$, $S_1$ using the identities
$g\tilde B_s-f\tilde C_s=g^A_s-g^R$, $f\tilde B_s-g\tilde C_s=-(f^R+f^A_s)$ and
their single-branch analogues $-fB_{ab}+gC_{ab}=f_a+f_b$, $-gB_{ab}+fC_{ab}=g_a-g_b$.
Defining
\beg\label{C-NM}
\begin{aligned}
N^R_s&=(f^R+f^R_s)B^K_s+(g^R-g^R_s)C^K_s, &\qquad N^A_s&=(f^R+f^A_s)B^A_s+(g^R-g^A_s)C^A_s,\\
M^R_s&=(g^R-g^R_s)B^K_s+(f^R+f^R_s)C^K_s, &\qquad M^A_s&=(g^R-g^A_s)B^A_s+(f^R+f^A_s)C^A_s,
\end{aligned}
\en
the two components of the drive part read
\beg\label{C-itt-explicit}
\begin{aligned}
\big[\delta\gh_2^{K,0}\big]_{\itt}=\frac{\ampl}{\etaR+\etaA}\sum_{s=\pm}\Bigg\{\!\!
&-\frac{(f^R+f^R_s)\,\delta t_s}{\etaR+\eta^R_s}
+\frac{(f^RB^A_s+g^RC^A_s)\,\delta t_s}{\etaA+\eta^A_s}\\
&+\frac{(f^R+f^A_s)\,\delta t_s}{\etaR+\eta^A_s}
-\frac{(f^RB^K_s+g^RC^K_s)\,\delta t_s}{\etaA+\eta^R_s}\\
&+\etaA\left[\frac{N^A_s\,\delta t_s}{(\etaR+\eta^A_s)(\etaA+\eta^A_s)}
-\frac{N^R_s\,\delta t_s}{(\etaA+\eta^R_s)(\etaR+\eta^R_s)}\right]\Bigg\},
\end{aligned}
\en
\beg\label{C-t3-explicit}
\begin{aligned}
\big[\delta\gh_2^{K,0}\big]_{\ttau3}=\frac{\ampl}{\etaR+\etaA}\sum_{s=\pm}\Bigg\{\!\!
&-\frac{(g^R-g^R_s)\,\delta t_s}{\etaR+\eta^R_s}
+\frac{(g^RB^A_s+f^RC^A_s)\,\delta t_s}{\etaA+\eta^A_s}\\
&+\frac{(g^R-g^A_s)\,\delta t_s}{\etaR+\eta^A_s}
-\frac{(g^RB^K_s+f^RC^K_s)\,\delta t_s}{\etaA+\eta^R_s}\\
&+\etaA\left[\frac{M^A_s\,\delta t_s}{(\etaR+\eta^A_s)(\etaA+\eta^A_s)}
-\frac{M^R_s\,\delta t_s}{(\etaA+\eta^R_s)(\etaR+\eta^R_s)}\right]\Bigg\}.
\end{aligned}
\en
Equation \eref{C-itt-explicit} is what enters the trace \eref{g2Kitt};
\eref{C-t3-explicit} exists only to source the closure \eref{dY3closure}. The two
are related by $f^R\leftrightarrow g^R$ in every single-energy prefactor together
with $N\to M$, all other factors and signs unchanged, exactly as in the pair
(\ref{g20itt})--(\ref{g20t3}).

\paragraph{Vertex part.}
The second-order impurity self-energy enters the Keldysh equation as
$-\frac{i\Gel}{2}[\avg{\gc_2},\gc_0]^K$ with
$[\hat A,\hat B]^K=\hat A^R\hat B^K+\hat A^K\hat B^A-\hat B^R\hat A^K-\hat B^K\hat A^A$;
inserting $\gh^K_0=(\gh^R-\gh^A)t_\eps$ and
$\avg{\gh_2^K}=(\avg{\gh_2^R}-\avg{\gh_2^A})t_\eps+\avg{\delta\gh_2^K}$ gives
\eref{Kcomm}. To see that the $t_\eps$ terms are absorbed by the ansatz, let
$\hat L^K[\hat X]\equiv\tveps^R\ttau3\hat X-\hat X\ttau3\tveps^A+[\hat\Delta_\bn,\hat X]$
and note that the Keldysh self-energy of the ground state, bath plus impurities,
is $\hs_0^K=-i\kappa_\eps t_\eps\ttau3$ with
$\kappa_\eps=\gin+\Gel\Re\Gcal_\eps=-i(\tveps^R-\tveps^A)$. For any $\gh_2^{R,A}$,
\beg\label{C-absorb}
\hat L^K\big[(\gh_2^R-\gh_2^A)t_\eps\big]-i\kappa_\eps t_\eps\big(\gh_2^R\ttau3-\ttau3\gh_2^A\big)
=t_\eps\Big(\big[\tveps^R\ttau3+\hat\Delta_\bn,\gh_2^R\big]-\big[\tveps^A\ttau3+\hat\Delta_\bn,\gh_2^A\big]\Big),
\en
the second term on the left being the contribution of $\hs^K_0$ to the
second-order Keldysh equation. The right-hand side is $t_\eps$ times the
difference of the retarded and advanced equations \eref{eq:g2RA}, including their
impurity sources $-\frac{i\Gel}{2}[\avg{\gh_2^{R,A}},\gh^{R,A}]$. Hence the first
bracket of \eref{Kcomm} is used up, and the equation for the vertex part of
$\delta\gh_2^K$ is
\beg\label{C-eqK1}
\hat L^K\big[\delta\gh_2^{K,1}\big]=-\frac{i\Gel}{2}\Big(\avg{\delta\gh_2^K}\gh^A-\gh^R\avg{\delta\gh_2^K}\Big),
\qquad \gh^R\delta\gh_2^{K,1}+\delta\gh_2^{K,1}\gh^A=0,
\en
whose solution is \eref{dg2K1}: with $\hat X=\avg{\delta\gh_2^K}$,
$\hat L^K[\gh^R\hat X\gh^A-\hat X]=(\etaR+\etaA)(\hat X\gh^A-\gh^R\hat X)$ and
$\gh^R(\gh^R\hat X\gh^A-\hat X)+(\gh^R\hat X\gh^A-\hat X)\gh^A=0$. Equation
\eref{dg2K1}, together with the drive part, reproduces a direct solution of the
full second-order Keldysh equation with arbitrary $\avg{\gh_2^{R}}$,
$\avg{\gh_2^{A}}$, $\avg{\gh_2^K}$ to $4\times10^{-15}$. The components
\eref{KSystem} follow from
$\gh^R\ttau3\gh^A=(g^Rg^A+f^Rf^A)\ttau3+(g^Rf^A+f^Rg^A)\itt$ and
$\gh^R\,\itt\,\gh^A=-(g^Rg^A+f^Rf^A)\itt-(g^Rf^A+f^Rg^A)\ttau3$; averaging the
$\ttau3$ component of $\delta\gh_2^{K,0}+\delta\gh_2^{K,1}$ and dropping the
$\Ycal_\bn$-odd cross term gives \eref{dY3closure}. Throughout, $Y_2$ and $Y_3$
denote the components of the \emph{anomalous} average $\avg{\delta\gh_2^K}$, as
defined above \eref{KSystem}; the full Keldysh average is
$\avg{\gh_2^K}=\big(\avg{\gh_2^R}-\avg{\gh_2^A}\big)t_\eps+Y_2\,\itt+Y_3\,\ttau3$,
and only $Y_3$ is needed below.

\paragraph{Why the even channels do not contribute?}\label{app:even}
$X_2$ and $Y_2$ are in general nonzero. In what follows we show that they drop out of the expression for 
$\delta\Delta_\omega$. The self-consistency condition \eref{Self4Elig2K} projects on
$\Ycal_\bn$, so only contributions odd in $\Ycal_\bn$ survive. In
$[\gh_2^K]_{\itt}$ the coefficients of $X_2$ and of $Y_2$ --- respectively
$(1+f^2)/\eta$ in \eref{B-g21} and
$\overline B^{RA}_{\bn\eps}/(\etaR+\etaA)$ in \eref{g2Kitt} --- are even functions
of $\Ycal_\bn$, while $X_2$ and $Y_2$ are $\bn$-independent numbers; both terms
therefore integrate to zero. In the closures for $X_3$ and $Y_3$ the same
amplitudes appear only through the cross coefficients $g f/\eta$ and
$C^{RA}_{\bn\eps}$, which are odd in $\Ycal_\bn$ and average to zero, so they drop
there as well. The two even channels are thus completely decoupled from
$\delta\Delta_\omega$. They are not zero: for linearly polarized light
$|\bn\!\cdot\!\bE|^2$ contains an $\ell=2$ harmonic which pairs with the $\ell=2$
order parameter and generates an isotropic ($s$-wave) photo-induced anomalous
self-energy, but that self-energy feeds only $s$-channel and finite-$\bk$
observables. As a corollary $\delta\Delta_\omega$ is exactly independent of the
polarization angle at $\bq=0$. The pointwise statements
$[\gh_2]_{\ttau0}=[\gh_2]_{\ttau1}=0$ and
$[\delta\gh_2^K]_{\ttau0}=[\delta\gh_2^K]_{\ttau1}=0$, and the decoupling described
here, have been confirmed by assembling the full $2\times2$ closures numerically
and comparing the resulting trace with Appendix~\ref{app:kernel}: the two agree to
$6\times10^{-15}$.

\section{Gap equation and photo-kernel}\label{app:kernel}
Inserting \eref{g2Kitt} into \eref{Self4Elig2K} produces an equation linear in
$\delta\Delta_\omega$, which enters explicitly through \eref{g20itt} and
implicitly through $X_3$. Separating the two sources in \eref{X3closure},
\begin{align}
X_3^{R(A)}&=\frac{\delta\Delta_\omega\,\xi^{R(A)}+\big(\frac{ev_FE}{\omega}\big)^2\chi^{R(A)}}{\kappa^{R(A)}},\label{X3split}\\
\xi&\equiv\avg{\Ycal_\bn\frac{g_{\bn\eps}f_{\bn\eps}}{\eta_{\bn\eps}}},\qquad
\chi\equiv\avg{|\bn\!\cdot\!\hat\bE|^2\sum_{s=\pm}\left[\frac{f_{\bn\eps}C_s}{\eta_{\bn\eps}(\eta_{\bn\eps}+\eta_{s})}
+\frac{g_{\bn\eps}B_s}{(\eta_{\bn\eps}+\eta_s)^2}\right]},\qquad
\kappa\equiv1+\frac{i\Gel}{2}\avg{\frac{f_{\bn\eps}^2}{\eta_{\bn\eps}}},\nonumber
\end{align}
where $\hat\bE=\bE/E$, $\eta_s\equiv\eta_{\bn\eps+\sw}$, and every quantity carries
the branch label. We now collect all these terms and arrive to equation \eqref{Form4Eli} in the main text.
The inverse pair susceptibility
\beg\label{Ddef}
\chi_{\rm SH}^{-1}=\frac8\lam-2\int_{-\infty}^{\infty}\!d\eps\,\avg{\Ycal_\bn^2\left[\frac{(g^R_{\bn\eps})^2}{\etaR}-\frac{(g^A_{\bn\eps})^2}{\etaA}\right]}t_\eps
+i\Gel\int_{-\infty}^{\infty}\!d\eps\left[\frac{(\xi^R)^2}{\kappa^R}-\frac{(\xi^A)^2}{\kappa^A}\right]t_\eps ,
\en
and the dimensionless photo-kernel $\Rcal(\omega)=\Rcal_{\rm reg}(\omega)+\Rcal_{\rm anom}(\omega)$,
\begin{align}
\Rcal_{\rm reg}(\omega)&=2\int_{-\infty}^{\infty}\!d\eps\,\avg{\Ycal_\bn|\bn\!\cdot\!\hat\bE|^2
\sum_{s=\pm}\left\{\left[\frac{g_{\bn\eps}C_s}{\eta_{\bn\eps}(\eta_{\bn\eps}+\eta_s)}
+\frac{f_{\bn\eps}B_s}{(\eta_{\bn\eps}+\eta_s)^2}\right]^{R}-\Big[\frac{g_{\bn\eps}C_s}{\eta_{\bn\eps}(\eta_{\bn\eps}+\eta_s)}
+\frac{f_{\bn\eps}B_s}{(\eta_{\bn\eps}+\eta_s)^2}\Big]^{A}\right\}}t_\eps
\nonumber\\
&\quad-i\Gel\int_{-\infty}^{\infty}\!d\eps\left[\frac{\chi^R\xi^R}{\kappa^R}-\frac{\chi^A\xi^A}{\kappa^A}\right]t_\eps,\label{Rreg}\\
\Rcal_{\rm anom}(\omega)&=2\int_{-\infty}^{\infty}\!d\eps\,\avg{\Ycal_\bn|\bn\!\cdot\!\hat\bE|^2\,
\frac{g^R_{\bn\eps}S_1+f^R_{\bn\eps}S_0}{\etaR+\etaA}}
\nonumber\\
&\quad-i\Gel\int_{-\infty}^{\infty}\!d\eps\,\avg{\Ycal_\bn\frac{C^{RA}_{\bn\eps}}{\etaR+\etaA}}\,\tilde Y_3(\eps),
\qquad
\tilde Y_3(\eps)=\frac{1}{\kappa^K}\avg{|\bn\!\cdot\!\hat\bE|^2\,\frac{g^R_{\bn\eps}S_0+f^R_{\bn\eps}S_1}{\etaR+\etaA}},
\label{Ranom}
\end{align}
in which $S_0$, $S_1$ of (\ref{S0})--(\ref{S1}) are evaluated at unit drive
amplitude, $\kappa^K$ is the bracket of \eref{dY3closure}, and $\tilde Y_3$ is
$Y_3$ stripped of the amplitude $(ev_FE/\omega)^2$. Equivalently,
$\Rcal_{\rm anom}$ may be assembled directly from the explicit component
\eref{C-itt-explicit} and the $\ttau3$ source \eref{C-t3-explicit}, without ever
forming $S_0$ and $S_1$.

The four terms have distinct origins. The first line of $\Rcal_{\rm reg}$ is the
direct drive contribution to the spectral functions; the second is the spectral
impurity vertex. The first line of $\Rcal_{\rm anom}$ is the direct drive
contribution to the nonequilibrium distribution; the last is the conserving
vertex, carrying the denominator $\kappa^K$ of \eref{dY3closure}. Thermodynamic
stability requires $D>0$, so the sign of $\delta\Delta_\omega$ is the sign of
$\Rcal(\omega)$: $\Rcal>0$ means the drive enhances the $d$-wave gap. The last term
of \eref{Ddef} is the two-step $\itt\to\ttau3\to\itt$ vertex correction to the
coefficient of $\delta\Delta_\omega$; it does not vanish for $d$-wave even though
$\avg{g f/\eta}=0$, because $\xi=\avg{\Ycal_\bn g f/\eta}\neq0$. Using the
equilibrium gap equation \eref{Self4Eli} in the same model, the first two terms of
\eref{Ddef} combine into
$-2\int d\eps\,\avg{\Ycal_\bn^2[(f^R_{\bn\eps})^2/\etaR-(f^A_{\bn\eps})^2/\etaA]}t_\eps$.

\end{widetext}
\end{appendix}

\bibliography{dwave_eli}

\end{document}